\documentclass[]{interact}

\usepackage{epstopdf}
\usepackage{subfigure}

\usepackage{amsmath}
\usepackage{amssymb}
\usepackage{amsthm}
\usepackage{dsfont}

\usepackage{graphicx}
\usepackage{booktabs}
\usepackage{multirow}
\usepackage{array}
\usepackage{longtable}

\usepackage{xcolor}
\usepackage{url}
\usepackage{tikz}
\usetikzlibrary{calc}

\usepackage[numbers,sort&compress,merge]{natbib}

\bibpunct[, ]{(}{)}{,}{n}{,}{,}

\theoremstyle{plain}
\newtheorem{theorem}{Theorem}[section]
\newtheorem{proposition}[theorem]{Proposition}

\theoremstyle{definition}

\theoremstyle{remark}

\begin{document}


\title{Intensity-Based Estimation of Extreme Value Models for Highly Censored Data via a Latent Poisson Process}

\author{
\name{
Eduardo Janotti\textsuperscript{a,b}\thanks{CONTACT Eduardo Janotti. Email: eduardojanotti@ime.usp.br},
Antonio Carlos Pedroso de Lima\textsuperscript{a},
L\'igia Henriques-Rodrigues\textsuperscript{b,c}
}
\affil{\textsuperscript{a}Departamento de Estat\'istica, Instituto de Matem\'atica, Estat\'istica e Ci\^encia da Computa\c{c}\~ao, Universidade de S\~ao Paulo, S\~ao Paulo, Brasil}
\affil{\textsuperscript{b}CEAUL -- Centro de Estat\'istica e Aplica\c{c}\~oes, Faculdade de Ci\^encias, Universidade de Lisboa, Lisboa, Portugal}
\affil{\textsuperscript{c}Departamento de Ci\^encias Matem\'aticas, Faculdade de Ci\^encias, Universidade de Lisboa, Lisboa, Portugal}
}

\maketitle

\begin{abstract}
Extreme value theory provides a natural framework for modelling rare events and extrapolating tail probabilities, but its direct application to survival analysis is complicated by right censoring, heterogeneous follow-up, and the need for coherent inference beyond the range of the observed data. We propose a Bayesian survival model that integrates extreme value theory with a hazard-based formulation built upon the extended generalized Pareto distribution. Rather than separating the bulk and tail of the distribution through an explicit threshold, the proposed model represents the entire hazard function within a unified parametric framework, allowing moderate and extreme survival times to be analysed simultaneously while preserving the asymptotic tail behaviour implied by extreme value theory. Inference is performed through a two-stage procedure. First, the hazard function is approximated by a piecewise exponential model. Second, these estimates are linked to the extended generalized Pareto distribution hazard through a hierarchical Bayesian model fitted using Hamiltonian Monte Carlo. We further develop bias-correction procedures based on latent representative locations and adaptive partition refinement, and establish theoretical results showing the asymptotic consistency of the proposed approximation under standard regularity conditions. Simulation studies demonstrate that the methodology provides reliable uncertainty quantification across a wide range of tail regimes, censoring proportions, and sample sizes, while remaining robust under moderate model misspecification. The practical performance of the approach is illustrated through an analysis of a heart failure cohort, where the proposed model enables inference on long-term survival probabilities and tail behaviour beyond the observed follow-up period. The resulting framework offers a computationally feasible and theoretically grounded alternative for modelling censored survival data when inference on extreme survival times is of primary scientific interest.
\end{abstract}



\section{Introduction}

The statistical modelling of rare and extreme events is fundamental to risk assessment and long-term prediction. Extreme Value Theory (EVT) provides a principled framework for characterizing distributional tails, with the Peaks-Over-Threshold (POT) approach and the Generalized Pareto (GP) distribution being among its principal methodologies \cite{Beirlant2004}. A central parameter in this framework is the extreme value index (EVI), denoted by $\xi$, which determines the asymptotic tail behaviour. Positive, zero, and negative values of $\xi$ correspond, respectively, to heavy-tailed distributions with no finite upper endpoint, the Gumbel domain characterized by exponential-type tail decay, and distributions with a finite upper endpoint. Estimation of $\xi$ is therefore essential for characterizing tail behaviour and extrapolating extreme probabilities.

Applying POT-based methods to survival data introduces additional challenges due to right censoring and heterogeneous follow-up times. In particular, information about the extreme survival tail may be limited to a small number of observed events and the censoring pattern, making extrapolation beyond the observed follow-up period difficult. Under substantial censoring, conventional likelihood-based estimation of GP and related tail models can become numerically unstable, with the tail-index estimate $\xi$ being especially sensitive to limited information. These difficulties are particularly relevant when the tail regime is unknown and $\xi$ must be estimated without imposing restrictions associated with a specific domain of attraction.

Several approaches have been developed to address extreme-value estimation under right censoring. Existing methods establish consistency and asymptotic properties for tail-index estimators in the presence of random censoring \cite{STUPFLER20161,10.1214/21-EJS1857,bladt2024censoredextremevalueestimation}. Related developments include cure-rate models for populations containing individuals who may never experience the event of interest \cite{Tsodikov,LopezSegovia2014}, as well as more flexible tail distributions such as the Extended Generalized Pareto (EGP) model, which allows additional flexibility in the body while preserving the asymptotic tail behaviour \cite{PAPASTATHOPOULOS2013131,CoolArticle,Extension}. Despite these developments, stable estimation of GP- and EGP-type models under severe censoring remains challenging, particularly when the tail index is unrestricted.

The present work is motivated by our previous study on extreme survival probabilities under right censoring \cite{janotti_novel}, which introduced a frequentist semi-parametric framework for incorporating extreme-value structure into survival analysis and extrapolating survival probabilities beyond observed follow-up times based on a latent intensity represented by a Log-Gaussian Cox Process (LGCP). That study highlighted computational and inferential difficulties associated with direct likelihood-based estimation in the extreme region. A subsequent work \cite{janotti2026beyond} considered a Bayesian extension. While this formulation demonstrated the potential of intensity-based modelling for extreme survival extrapolation, the latent Gaussian-process structure introduces additional modelling and computational complexity, including the numerical treatment of the latent intensity. Related challenges in Gaussian-process maximum likelihood estimation have also been documented in the literature \cite{karvonen2023maximumlikelihoodestimationgaussian,Basak_2022}.

In this paper, we propose an alternative estimation procedure for GP and EGP tail models based on a latent Poisson-process representation of the intensity associated with the underlying extreme-value model. Unlike the previous LGCP formulation, the proposed approach does not introduce an additional latent Gaussian process, yielding a simpler probabilistic structure and an alternative computational strategy for tail-parameter estimation. The method is designed to address numerical difficulties arising under substantial censoring, with particular emphasis on estimating the tail index $\xi$ without restricting it a priori to a specific tail regime. This allows the data to inform whether the underlying tail is heavy, of Gumbel type, or has a finite upper endpoint.

The proposed framework further provides a basis for extensions to covariate-dependent tail behaviour and bias-correction procedures. These developments are intended to improve the practical use of extreme-value models for long-term survival extrapolation when the available event information is sparse. The methodology is evaluated through simulation studies examining finite-sample performance, numerical stability, and robustness, and is illustrated using a heart failure dataset.

The remainder of the paper is organized as follows. Section~\ref{sec:state_art} reviews the relevant background on EVT, GP and EGP models, and intensity-based representations for censored survival data. Section~\ref{sec:MYMODEL} introduces the proposed latent Poisson-process formulation and Bayesian estimation procedure. Section~\ref{sec:LookAtMyTail} develops extensions for covariate-dependent tail behaviour, while Section~\ref{sec:bias_correction} discusses bias correction based on latent representative locations and adaptive partition refinement. Section~\ref{sec:simulation} presents the simulation studies, and Section~\ref{sec:application} provides the heart failure data application. Finally, Section~\ref{sec:conclusions} summarizes the findings, limitations, and directions for future research.

\section{State of the art and recent developments}
\label{sec:state_art}

\subsection{Fundamentals}

EVT provides the asymptotic foundations for statistical methods designed to characterize distribution tails and quantify the probability of extreme events. Its main results underpin widely used approaches for modelling extremes, including block maxima and threshold exceedances, and motivate the GP- and EGP-based models considered in this work.

The first cornerstone result is the Fisher--Tippett--Gnedenko theorem \cite{FisherTippett1928,Gnedenko1943}, which states that, under suitable conditions and normalization, the maximum of a sequence of independent and identically distributed random variables converges in distribution to a non-degenerate limit belonging to the Generalized Extreme Value (GEV) family. This theorem establishes the theoretical basis for block-maxima methods and characterizes the possible limiting distributions of normalized maxima.

While block-maxima methods derived from this theorem are theoretically appealing, they can be inefficient in practice because only a subset of the available observations contributes to the analysis. Threshold-based approaches address this limitation through the second fundamental result in EVT, the Pickands--Balkema--de Haan theorem \cite{Balkema1974,Pickands1975}. Under suitable regularity conditions, the same as the previous result, this theorem states that the conditional distribution of exceedances above a sufficiently high threshold can be approximated by a GP. Specifically, if $F$ denotes the cumulative density function (CDF) of a random variable $X$ with right endpoint
\(
x_F=\sup\{x\in\mathbb{R}:F(x)<1\},
\)
the conditional excess CDF above a threshold $u<x_F$ is defined by
\begin{equation}
  P(X-u\leq y\mid X>u)
=
\frac{F(u+y)-F(u)}{1-F(u)},
\qquad
0\leq y<x_F-u.  
\end{equation}
For sufficiently large $u$, this distribution is well approximated by a GP with scale parameter $\sigma>0$ and shape parameter $\xi$, where $\xi$ is the same EVI as in the Fisher--Tippett--Gnedenko theorem. Since the GP provides the only possible non-degenerate limiting family for threshold exceedances under the theorem's conditions, it occupies a central position in POT methodology.

Comprehensive treatments of parametric and semi-parametric inference for GEV and GP models can be found in \cite{coles2001introduction} and \cite{Beirlant2004}. While the former emphasizes practical modelling strategies and diagnostic tools, the latter develops a broad range of estimation techniques for tail-related parameters, including semi-parametric approaches.

A persistent challenge in POT modelling is the estimation of the EVI. Likelihood-based inference for the GP can present non-standard behaviour, including parameter-dependent constraints and difficulties near the boundary of the parameter space, while higher-order moments may fail to exist for certain values of $\xi$. Classical parametric approaches include maximum likelihood and probability-weighted moments \cite{BermudezKotz2010}, while alternative procedures such as the elemental percentile estimator \cite{CastilloHadi1997} provide direct approaches to tail-index estimation, in some cases under restricted parameter ranges. In semi-parametric settings, moment-based, Pickands, and Hill-type estimators are also widely employed, with their performance depending on the underlying tail behaviour and on the amount of information available in the extreme region \cite{Beirlant2004}. These issues become particularly important when the number of extreme observations is small or when the data are subject to substantial censoring. An additional fundamental challenge is the choice of the threshold above which the POT model is fitted, which remains an open problem in extreme value analysis. A threshold that is too low may introduce bias by violating the asymptotic approximation underlying the GP model, whereas a threshold that is too high reduces the number of exceedances available for inference, increasing estimation uncertainty. Consequently, selecting an appropriate threshold requires balancing bias and variance, and this trade-off becomes especially challenging in the presence of limited extreme observations or substantial censoring.

In parallel with EVT, survival analysis focuses on modelling time-to-event data, typically in the presence of censoring. Right-censoring occurs when the event of interest is not observed during the follow-up period, so that the failure time is only known to exceed the censoring time. This mechanism is ubiquitous in medical, reliability, and demographic studies. A particularly important setting is random right-censoring, in which the censoring time follows its own stochastic mechanism and is assumed to be independent of the failure time, possibly conditional on observed covariates. For more details, see \cite{Aalen2008}.

Let $X$ denote a nonnegative random variable representing the failure time, with density $f$ and survival function
\(
S(x):=P(X>x).
\)
The hazard function is defined by
\(
h(x):=f(x)/S(x),
\)
whenever $S(x)>0$.

When censoring is present, the observed time and censoring indicator are
\begin{equation}
    T=\min\{X,C\},
\qquad
\Delta=\mathbb{I}\{X\leq C\},
\end{equation}
where \(C\) denotes the censoring time. Under independent censoring, \(C\) is assumed to be independent of \(X\); more generally, conditional independence may be assumed given observed covariates.

Let \((T_i,\Delta_i)_{i=1}^n\) be independent and identically distributed copies of \((T,\Delta)\), and let \((t_i,\delta_i)\) denote their observed realizations. The observed follow-up data are therefore \((t_i,\delta_i)_{i=1}^n\), where
\begin{equation}
    \delta_i=
\begin{cases}
1, & \text{if the event is observed at } t_i,\\
0, & \text{if the observation is right-censored at } t_i.
\end{cases}
\end{equation}
Given these observed pairs, the primary inferential target is often the survival function \(S(x)\) or functionals derived from it. Classical nonparametric estimators, such as the Kaplan--Meier and Nelson--Aalen estimators, are widely used because of their robustness and relatively weak modelling assumptions \cite{andersen1993statistical,fleming2005counting}. However, their ability to provide information about the survival function becomes increasingly limited in the extreme tail. In particular, once the largest observed failure time has been reached, nonparametric estimators do not provide a principled mechanism for extrapolating the survival function to times beyond the observed range. Consequently, estimation of long-term survival probabilities requires additional assumptions or the specification of a parametric or semi-parametric tail model.

These limitations highlight a fundamental disconnect between standard survival-analysis tools and the objectives of extreme-tail inference. Most conventional survival models are not specifically designed to estimate the EVI or to distinguish between heavy-, exponential- and bounded-tail behaviour. Conversely, many classical EVT applications focus on heavy-tailed settings, whereas survival data may exhibit a much broader range of tail behaviours. Biomedical and demographic processes, for example, may be compatible with rapidly decaying tails or with distributions having an effective or finite upper endpoint. Consequently, there is a need for methodological frameworks that integrate EVT principles with survival analysis while allowing the data to inform the relevant tail regime.

This issue is particularly challenging under substantial right-censoring. When few failures are observed in the extreme region, direct likelihood-based estimation of GP parameters may become numerically unstable, and estimation of the EVI can be highly sensitive to the available tail information. In such situations, imposing a priori restrictions on $\xi$ to distinguish between specific tail regimes may facilitate computation but can also constrain the inferential problem. An estimation framework that can accommodate different values of $\xi$ without requiring such regime-specific restrictions is therefore of particular interest for long-term survival extrapolation.

These considerations motivate the developments pursued in this paper. We focus on an alternative intensity-based estimation procedure for GP and EGP models, in which the relevant tail observations are represented through a latent Poisson-process structure. The resulting formulation provides a different computational route to inference on the tail parameters and is designed to improve numerical stability in highly censored settings, with particular emphasis on the estimation of the EVI.

\subsection{Some classical alternatives}

A substantial work on long-term survival modelling is built around the notion that a non-negligible fraction of the population may never experience the event of interest. \cite{LopezSegovia2014} provides a thorough review of such models, with particular attention to cure-rate formulations. In this class of models, the population is typically represented as a mixture of susceptible and cured individuals, with the latter characterized by a positive limiting survival probability. The limiting value of the survival function is commonly referred to as the cure fraction, and its estimation plays a central role in long-term survival inference.

From a complementary perspective, several authors have focused on extreme-value inference under right random censoring without explicitly invoking a cure mechanism. In this line of research, the primary objective is the estimation of tail-related quantities, as the EVI and high quantiles, when observations are subject to stochastic censoring. Representative contributions include the incorporation of covariates into GP-based estimators \cite{STUPFLER20161}, locally robust procedures \cite{Dierckx2021}, trimming-based estimators for heavy-tailed data \cite{10.1214/21-EJS1857}, and conditional approaches motivated by economic applications \cite{https://doi.org/10.1111/ecin.13119}. More recently, a general framework for censored extreme-value estimation was developed in \cite{bladt2024censoredextremevalueestimation,bladt2024censoredextremelossesfunctional}, extending generalized Hill-type estimators, establishing asymptotic theory, proposing diagnostic and goodness-of-fit tools.

Another classical strategy for retaining information from the entire sample consists of combining separate models for the bulk and the tail of the distribution. In its simplest form, this approach specifies one model below a threshold and a POT-based model above it. In practice, however, a valid hybrid density generally requires appropriate normalization and, when continuity is desired, matching conditions at the threshold. The basic construction can nevertheless be interpreted as a hybrid model in which EVT tools are reserved for the upper tail, while the central part of the data is handled through a conventional survival or regression model. Such formulations are closely related to splicing methods, in which different distributions are combined to represent different regions of the data distribution \cite{klugman2019loss}.

Within survival analysis, this idea was explored by Grama et al.~\cite{grama:hal-00918822}, who modelled exceedances above a threshold using an exponential distribution, corresponding to a GP with $\xi=0$, while incorporating lower observations through the Kaplan--Meier estimator. Subsequent extensions \cite{Grama04072019,YU2023109880} addressed heterogeneity by combining POT modelling for extremes with Cox-type regression for the bulk of the data, estimated using the Breslow estimator \cite{Lin2007} and further refined through penalized likelihood techniques.

While hybrid and slicing-based approaches offer considerable flexibility, they inherit the intrinsic difficulty of threshold selection. The choice of threshold can strongly influence the bias and variance of tail estimates and remains one of the most delicate practical aspects of POT modelling. Numerous data-driven selection rules and alternative threshold-selection procedures have been proposed; a comprehensive overview is given in \cite{belzile2026choosingthresholdextremevalue}.

Building on this line of research, several authors have developed threshold-free constructions within the EGP framework. In particular, \cite{CoolArticle} proposed a formulation that preserves GP tail behaviour while allowing the central part of the distribution to vary flexibly through a functional composition involving the GP CDF and a constrained perturbation CDF. This approach avoids explicit threshold selection and exploits the full dataset while maintaining asymptotic tail coherence. Although the original formulation introduced several possible perturbation mechanisms, no formal selection criterion was provided, and the specification remained largely pragmatic.

Subsequent contributions expanded the EGP paradigm. For example, \cite{Stein2021a,Stein2021b,https://doi.org/10.1002/env.2744,LI2023107761} proposed alternative constructions, while \cite{carrer2022distributionalregressionmodelsextended} incorporated EGP-based structures into distributional regression models. Furthermore, \cite{Extension} extended the framework to a semi-parametric setting by estimating the perturbation density using Bernstein polynomials rather than fixing it \emph{a priori}.

Two complementary challenges emerge from the existing literature. On the one hand, censored EVT methods provide principled tools for estimating extreme-value parameters under incomplete observation but may rely on estimators whose numerical behaviour deteriorates when the available tail information is sparse. On the other hand, threshold-free EGP constructions allow the bulk and tail of the distribution to be modelled within a unified framework, but their existing estimation procedures do not directly address the computational difficulties that arise in highly censored settings.

\section{The proposed model}
\label{sec:MYMODEL}

We propose a Bayesian survival model that embeds the tail behaviour of the EGP distribution directly into the hazard function. Rather than explicitly separating threshold exceedances from the bulk of the observations, the model specifies the hazard over the entire time domain. The parameters govern the overall hazard shape, while the EVI, $\xi$, determines its asymptotic tail behaviour. Bayesian inference allows prior information to be incorporated and posterior uncertainty to be quantified.

The model is formulated within a counting-process framework, which provides a natural representation of time-to-event data and connects hazard-based and intensity-based modelling \cite{Aalen2008}. Let \(T_i=\min(X_i,C_i)\) and \(\Delta_i=\mathbb{I}(X_i\leq C_i)\) denote the random observed follow-up time and event indicator for individual \(i\), respectively. For each individual, define the counting process
\begin{equation}
N_i(t)=\mathbb{I}(T_i\leq t,\Delta_i=1),
\qquad t\geq 0,
\end{equation}
which records whether the event has occurred by time \(t\), and the at-risk process
\begin{equation}
Y_i(t)=\mathbb{I}(T_i\geq t),
\qquad t\geq 0.
\end{equation}
The counting process has intensity
\begin{equation}
\lambda_i(t)=Y_i(t)h(t),
\end{equation}
where \(h(t)\) is the underlying hazard function. Its Doob--Meyer decomposition is
\begin{equation}
N_i(t)
=
\int_0^t Y_i(u)h(u)\,\mathrm{d}u+M_i(t),
\end{equation}
where \(M_i(t)\) is a local martingale with respect to the underlying filtration.

Building on this representation, we combine extreme value theory with an intensity-based latent Poisson-process formulation to construct an alternative estimation procedure for GP and EGP models. By exploiting information from censored observations, this approach aims to improve the stability of EVI estimation across different tail regimes, while preserving the asymptotic properties implied by the underlying extreme-value model.

\subsection{Extended generalized Pareto hazard}

Let
\begin{equation}
G(t;\sigma,\xi)
=
1-
\left(
1+\frac{\xi t}{\sigma}
\right)^{-1/\xi},
\qquad
1+\frac{\xi t}{\sigma}>0, \qquad t > 0,
\end{equation}
denote the CDF of a GP distribution, where \(\sigma>0\) is the scale parameter and \(\xi\) is the EVI. The corresponding density is
\begin{equation}
g(t;\sigma,\xi)
=
\frac{1}{\sigma}
\left(
1+\frac{\xi t}{\sigma}
\right)^{-1/\xi-1}.
\end{equation}
For convenience, define
\begin{equation}
q(t)
=
\left(
1+\frac{\xi t}{\sigma}
\right)^{-1/\xi}.
\end{equation}
Then,
$
G(t;\sigma,\xi)=1-q(t)
$
and
$
g(t;\sigma,\xi)
=
\frac{1}{\sigma}q(t)^{1+\xi}.
$

The EGP distribution is obtained by applying a power transformation to the GP CDF, which is defined as

\begin{equation}
G_{\mathrm{EGP}}(t;\sigma,\xi,\kappa)
=
\left\{G(t;\sigma,\xi)\right\}^{\kappa},
\qquad
\kappa>0,
\end{equation}
where \(\kappa\) is an additional shape parameter that controls the departure of the EGP distribution from the standard GP model. In particular, \(\kappa=1\) recovers the GP distribution, whereas values of \(\kappa\) different from one modify the distribution of observations away from the asymptotic tail.

Differentiating the EGP CDF with respect to \(t\) gives the corresponding density,
\begin{equation}
g_{\mathrm{EGP}}(t;\sigma,\xi,\kappa)
=
\kappa
\left\{G(t;\sigma,\xi)\right\}^{\kappa-1}
g(t;\sigma,\xi).
\end{equation}
Consequently, the EGP hazard function is
\begin{equation}
h_{\mathrm{EGP}}(t;\sigma,\xi,\kappa)
=
\frac{
\kappa
\left\{G(t;\sigma,\xi)\right\}^{\kappa-1}
g(t;\sigma,\xi)
}{
1-\left\{G(t;\sigma,\xi)\right\}^{\kappa}
}.
\end{equation}
Substituting \(G(t;\sigma,\xi)=1-q(t)\) and the expression for the GP density yields
\begin{equation} \label{eq:hazard_egp}
h_{\mathrm{EGP}}(t;\sigma,\xi,\kappa)
=
\frac{\kappa}{\sigma}
\frac{
[1-q(t)]^{\kappa-1}q(t)^{1+\xi}
}{
1-[1-q(t)]^\kappa
}.
\end{equation}

The parameter \(\kappa\) determines how the EGP distribution departs from the GP model in the body of the distribution. When \(0<\kappa<1\), the power transformation increases \(G(t;\sigma,\xi)^\kappa\) relative to \(G(t;\sigma,\xi)\), resulting in greater cumulative probability at a given finite value of \(t\). Conversely, when \(\kappa>1\), the transformation decreases \(G(t;\sigma,\xi)^\kappa\), resulting in less cumulative probability at the same value of \(t\). Thus, \(\kappa<1\) and \(\kappa>1\) provide two directions of departure from the GP model, while \(\kappa=1\) corresponds exactly to the standard GP distribution.

Importantly, the power transformation does not alter the first-order tail behaviour determined by the extreme value index \(\xi\). Indeed, as \(t\) approaches the upper tail, \(G(t;\sigma,\xi)\) approaches one, and the EGP survival function satisfies
\begin{equation}
1-G_{\mathrm{EGP}}(t;\sigma,\xi,\kappa)
=
1-\left\{G(t;\sigma,\xi)\right\}^{\kappa}
\sim
\kappa\left[1-G(t;\sigma,\xi)\right].
\end{equation}
Hence, the EGP distribution retains the same extreme value index \(\xi\) as the underlying GP distribution, while \(\kappa\) provides additional flexibility in modelling the distribution away from the asymptotic tail.

\subsection{Two-stage inference}
\label{subsec:2-stage}

Direct likelihood-based inference for GP models presents difficulties across different tail regimes. In particular, the regularity conditions underlying standard maximum likelihood estimation fail when $\xi<-1/2$, complicating inference for sufficiently negative values of the extreme value index. Conversely, moment-based estimation relies on the existence of the corresponding moments, which imposes restrictions on $\xi$ and limits its applicability for sufficiently heavy-tailed distributions. These difficulties are further aggravated by right censoring, which reduces the information available in the extreme region and makes likelihood evaluation and optimization more challenging. Bayesian procedures provide an alternative, but their computational cost can become substantial for large samples.

To address these limitations, we propose a two-stage approximation that avoids direct evaluation of the likelihood associated with the continuous-time EGP model. Consequently, the procedure does not require a closed-form expression for this likelihood. In the first stage, the censored survival observations are approximated through a piecewise exponential model, yielding interval-specific hazard estimates and their associated uncertainties. This provides a simpler representation of the original data while retaining information about both the estimated hazards and their precision.

In the second stage, these interval-specific estimates are related to the continuous-time EGP model through latent representative locations. The EGP parameters, including the extreme value index $\xi$, are then estimated from this approximate hazard representation rather than directly from the original censored-data likelihood. This construction separates the estimation of local hazard information from the estimation of the global tail structure, providing an alternative computational strategy for inference under substantial censoring.

The theoretical properties of the proposed construction are established in Appendix~A.

\subsubsection*{Step 1: Piecewise exponential approximation}

Let $
\mathcal{T}=(\rho_1,\rho_{M+1}]\subset\mathbb{R}^+
$
denote the time domain over which the hazard function is to be approximated. Consider a sequence of ordered partition points
$
\rho_1<\rho_2<\cdots<\rho_{M+1},
$
which divides \(\mathcal{T}\) into \(M\) consecutive intervals,
$
I_j=(\rho_j,\rho_{j+1}]$,
$
j=1,\ldots,M.
$
The hazard function is assumed to be constant within each interval \(I_j\), yielding the piecewise exponential approximation.
The specific procedure used to determine the partition points, including the criterion for dividing the time domain into intervals with similar event patterns, is described in Section~5.2.

Within each interval $I_j$, we approximate the hazard by a constant,
\begin{equation}
h(t)=h_j,
\qquad t\in I_j,
\end{equation}
with $h_j>0$. Under this assumption, conditional on being at risk at the beginning of $I_j$, the event process within the interval is a homogeneous Poisson process with rate $h_j$. Equivalently, by the memoryless property of the exponential distribution, the waiting time from $\rho_j$ to an event, conditional on survival to $\rho_j$, is

\begin{equation}
P(X>\rho_j+w\mid X>\rho_j)=\exp(-h_jw),
\qquad 0\leq w\leq \rho_{j+1}-\rho_j,
\end{equation}
assuming that the hazard of $X$ is constant only within interval \(I_j\). An event occurring within $I_j$ is therefore observed at its corresponding waiting time, whereas an individual whose observation terminates before an event occurs contributes a right-censored exponential observation at the end of their available follow-up in the interval.

To formalize these contributions, define the exposure time contributed by individual $i$ to interval $I_j$ as
\begin{equation}
E_{ij}
:=
\int_{I_j}Y_i(t)\,dt
=
\left[\min(t_i,\rho_{j+1})-\rho_j\right]_+,
\end{equation}
where $[x]_+=\max(x,0)$. Thus, $E_{ij}$ is the amount of time for which individual $i$ is under observation and at risk within $I_j$.

Similarly, define the interval-specific event indicator $
D_{ij}
=
\delta_i\, \mathbb{I}(t_i\in I_j).
$
Hence, $D_{ij}=1$ if individual $i$ experiences the observed event in interval $I_j$, and $D_{ij}=0$ otherwise. Since each individual can experience at most one event,
$
\sum_{j=1}^M D_{ij}\leq 1.
$
The likelihood contribution of individual $i$ in interval $I_j$ is
\begin{equation}
L_{ij}(h_j)
=
h_j^{D_{ij}}\exp\left(-h_jE_{ij}\right).
\end{equation}
The factor \(h_j^{D_{ij}}\) accounts for an observed event, while the exponential term accounts for the survival or right-censoring contribution over the corresponding exposure time. Aggregating over individuals, define
\begin{equation}
\begin{aligned}
D_j &:= \sum_{i=1}^n D_{ij},
&\qquad
E_j &:= \sum_{i=1}^n E_{ij},
\end{aligned}
\end{equation}
where \(D_j\) is the total number of observed events and \(E_j\) is the total exposure time accumulated by all individuals in interval \(I_j\).

For interval \(I_j\), the log-likelihood is
\begin{equation}
\ell_j(h_j)=D_j\log(h_j)-h_jE_j,
\end{equation}
yielding the maximum-likelihood estimate
\begin{equation}
\widehat h_j=\frac{D_j}{E_j},
\end{equation}
and the corresponding estimated standard error
\begin{equation}
\widehat{\tau}_j
=\mathcal{I}_j(\widehat h_j)^{-1/2}
=\frac{\widehat h_j}{\sqrt{D_j}}
=\frac{\sqrt{D_j}}{E_j},
\end{equation}
where \(\mathcal{I}_j(\widehat h_j)\) denotes the observed Fisher information evaluated at the maximum-likelihood estimate.
Thus, for each interval with \(D_j>0\) and \(E_j>0\), the piecewise exponential approximation is characterized by \((\widehat h_j,\widehat{\tau}_j)\), with estimated hazard function
\begin{equation}
\widehat h(t)
=\sum_{j=1}^M\widehat h_j\,\mathbb{I}(t\in I_j).
\end{equation}
Here, $\widehat h_j$ estimates the constant hazard assigned to interval $I_j$, while $\widehat{\tau}_j$ quantifies the corresponding estimation uncertainty. In particular, the standard error reflects the amount of event information available in each interval, with fewer observed events generally leading to greater estimation uncertainty. This distinction is important when the data are sparse or heavily censored, since the resulting hazard estimates may be less precise.

Intervals with $D_j=0$ and $E_j>0$ require special consideration. In this case, the likelihood is maximized at the boundary, yielding $\widehat h_j=0$, rather than a positive interior estimate. However, the observed-information-based standard error is not defined at this boundary estimate. Consequently, such intervals are excluded from the subsequent hazard-location procedure when that procedure requires a positive hazard estimate and a finite standard error.

\begin{figure}[htbp]
    \centering
    \includegraphics[
        width=0.85\textwidth
    ]{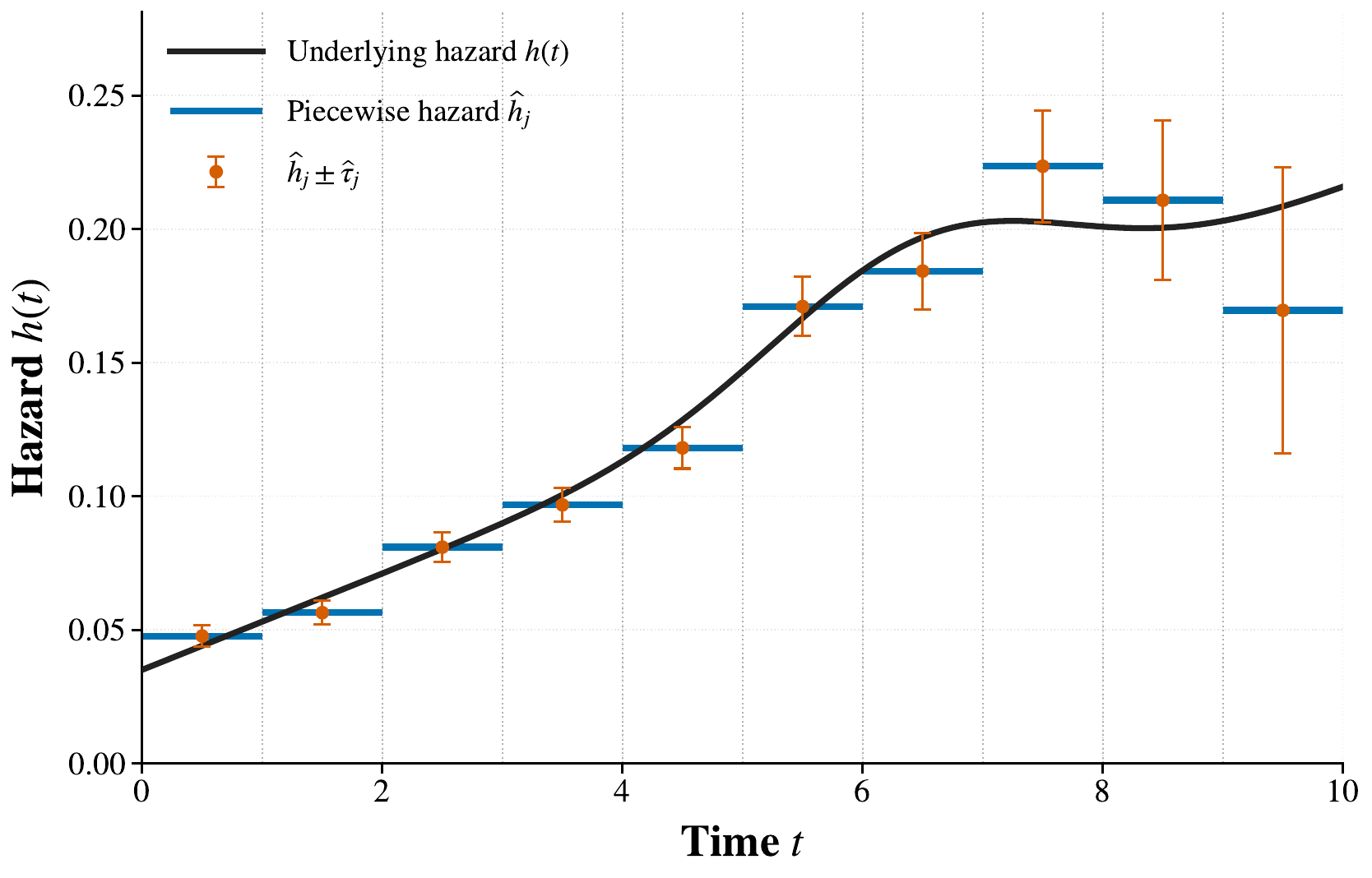}
    \caption{
    Illustration of the piecewise exponential approximation with
    $M=10$ intervals and no censoring.
    The black curve represents the underlying smooth hazard $h(t)$,
    while the blue step function gives the interval-specific
    maximum likelihood estimates $\widehat{h}_j = D_j/E_j$.
    The orange points and error bars represent
    $\widehat{h}_j \pm \widehat{\tau}_j$, where
    $\widehat{\tau}_j = \widehat{h}_j/\sqrt{D_j}$.
    Vertical dotted lines indicate the boundaries of the partition
    $\mathcal{T} = [\rho_1,\rho_{M+1}]$.
    }
    \label{fig:piecewise_exponential}
\end{figure}

Because the underlying EGP hazard is generally not constant within $I_j$, $\widehat h_j$ cannot, in general, be associated with a unique deterministic time point. We therefore associate each interval with a latent representative location
\begin{equation}
L_j\in I_j,
\qquad
\mathbb{P}(L_j\in I_j)=1.
\end{equation}
The choice of the latent locations is described in detail in Section~5.3. Figure~\ref{fig:piecewise_exponential} illustrates the piecewise exponential approximation obtained by partitioning the time domain into intervals $I_j=(\rho_j,\rho_{j+1}]$. The smooth underlying hazard is approximated within each interval by the constant estimate $\widehat{h}_j=D_j/E_j$, with uncertainty quantified by $\widehat{\tau}_j=\widehat{h}_j/\sqrt{D_j}$.

\subsubsection*{Step 2: EGP hazard model}

The first step provides, for each interval $I_j$, a piecewise exponential hazard estimate $\widehat h_j$ together with its estimated standard error $\widehat{\tau}_j$. Since $\widehat h_j$ summarizes the hazard over an interval rather than at a uniquely defined time point, we use the latent representative location $L_j\in I_j$ introduced above to connect the interval-level estimates with the underlying continuous-time EGP hazard.

Conditional on the latent location $L_j$, we model the first-stage hazard estimate using the hazard function implied by the EGP model in \eqref{eq:hazard_egp}, as follows
\begin{equation}
\widehat h_j
\mid L_j,\sigma,\xi,\kappa
\sim
\mathcal{N}^{+}
\left(
h_{\mathrm{EGP}}(L_j;\sigma,\xi,\kappa),
\widehat{\tau}_j^2
\right),
\end{equation}
where \(\mathcal{N}^{+}(\mu,\tau^2)\) denotes a normal distribution with location parameter \(\mu\) and scale parameter \(\tau^2\), truncated to the positive real line. Note that, due to the truncation, \(\mu\) and \(\tau^2\) do not generally coincide with the mean and variance of the resulting distribution. The location parameter is specified by the EGP hazard evaluated at the latent location \(L_j\), while the scale parameter is determined by the estimated uncertainty from the piecewise exponential approximation. Thus, the second stage does not treat \(\widehat h_j\) as an exact observation of the underlying hazard; instead, it explicitly propagates the estimation uncertainty from Step 1 through \(\widehat{\tau}_j^2\).

We model the hazard itself, rather than its logarithm, so that the regression remains on the original intensity scale. Positivity is enforced through the truncated normal distribution, which is consistent with the non-negative nature of the hazard.

Combining the two stages, the resulting model can be viewed as a latent-location regression of the piecewise exponential hazard estimates onto the continuous EGP hazard function. The first stage reduces the censored survival data to interval-specific hazard estimates and their associated uncertainties, while the second stage uses the latent locations to map these estimates back to the continuous time scale and estimate the EGP parameters $(\sigma,\xi,\kappa)$. The procedure for specifying the latent locations $L_j$ is given in Section~5.3.

\subsection{Prior specification and posterior inference}
\label{priors_for_everyone}

All model parameters are assumed to be \emph{a priori} independent. As a weakly informative default specification, the priors for the scale parameter $\sigma$ and the EGP shape parameter $\kappa$ are
\begin{equation}
\sigma \sim \operatorname{Lognormal}(0,10),
\qquad
\kappa \sim \operatorname{Lognormal}(0,10).
\end{equation}
These specifications may be adapted to the application and the available prior information, and are not essential to the parameterization of the extreme value index $\xi$ described below.

The support of the GP distribution imposes a constraint on $\xi$. Since the model is defined over the time domain $\mathcal{T}$, we require
\begin{equation}
1+\frac{\xi t}{\sigma}>0,
\qquad \forall\,t\in\mathcal{T}.
\end{equation}
Assuming that $\mathcal{T}$ has a finite, positive supremum, this condition is equivalent to
\begin{equation}
\label{eq:condition}
\xi>-\frac{\sigma}{\sup\mathcal{T}}.
\end{equation}
Rather than imposing this constraint directly, we introduce an unconstrained latent parameter with a centered Laplace prior,
\begin{equation}
\zeta\sim\operatorname{Laplace}(0,1),
\end{equation}
and define the extreme value index through the transformation
\begin{equation}
\label{eq:xi_transformation}
\xi
=
\frac{\sigma}{\sup\mathcal{T}}
\left(e^\zeta-1\right).
\end{equation}
For every finite value of $\zeta$, this transformation guarantees that inequality~\eqref{eq:condition} holds, while allowing $\xi$ to take arbitrarily large positive values. It therefore provides an unconstrained parameterization of the entire admissible range of $\xi$ over the specified time domain.

To characterize the induced prior on $\xi$, let
\begin{equation}
\label{eq:a_definition}
a=\frac{\sigma}{\sup\mathcal{T}},
\end{equation}
so that
\begin{equation}
\xi=a(e^\zeta-1).
\end{equation}
Because this transformation is strictly increasing and the Laplace distribution is centered at zero, the median of the induced prior is
\begin{equation}
\label{eq:xi_median}
\operatorname{median}(\xi\mid\sigma)=a(e^0-1)=0.
\end{equation}
Thus, the prior is centered at the exponential-tail boundary rather than favouring either a bounded-tail or a heavy-tailed regime a priori.

The induced density follows from a change of variables. Specifically, the inverse transformation and its derivative are
\begin{equation}
\zeta=\log\left(1+\frac{\xi}{a}\right),
\qquad
\left|\frac{\mathrm{d}\zeta}{\mathrm{d}\xi}\right|
=\frac{1}{a+\xi}.
\end{equation}
Consequently, the conditional density of $\xi$ is
\begin{equation}
\label{eq:induced_prior_xi}
p(\xi\mid\sigma)
=
\begin{cases}
\dfrac{1}{2a},
& -a<\xi<0,\\[6pt]
\dfrac{a}{2(a+\xi)^2},
& \xi\geq 0,\\[6pt]
0,
& \text{otherwise}.
\end{cases}
\end{equation}
The negative component is therefore uniform on the interval $(-a,0)$, conditional on $\xi<0$, and is bounded below by the GP support constraint. The positive component is unbounded and follows a Lomax (Pareto Type II) distribution, allowing for increasingly heavy-tailed behaviour. The resulting prior is asymmetric in both its support and density, despite being centered at zero in the median sense.

\begin{figure}[htbp]
    \centering
    \includegraphics[width=0.85\textwidth]{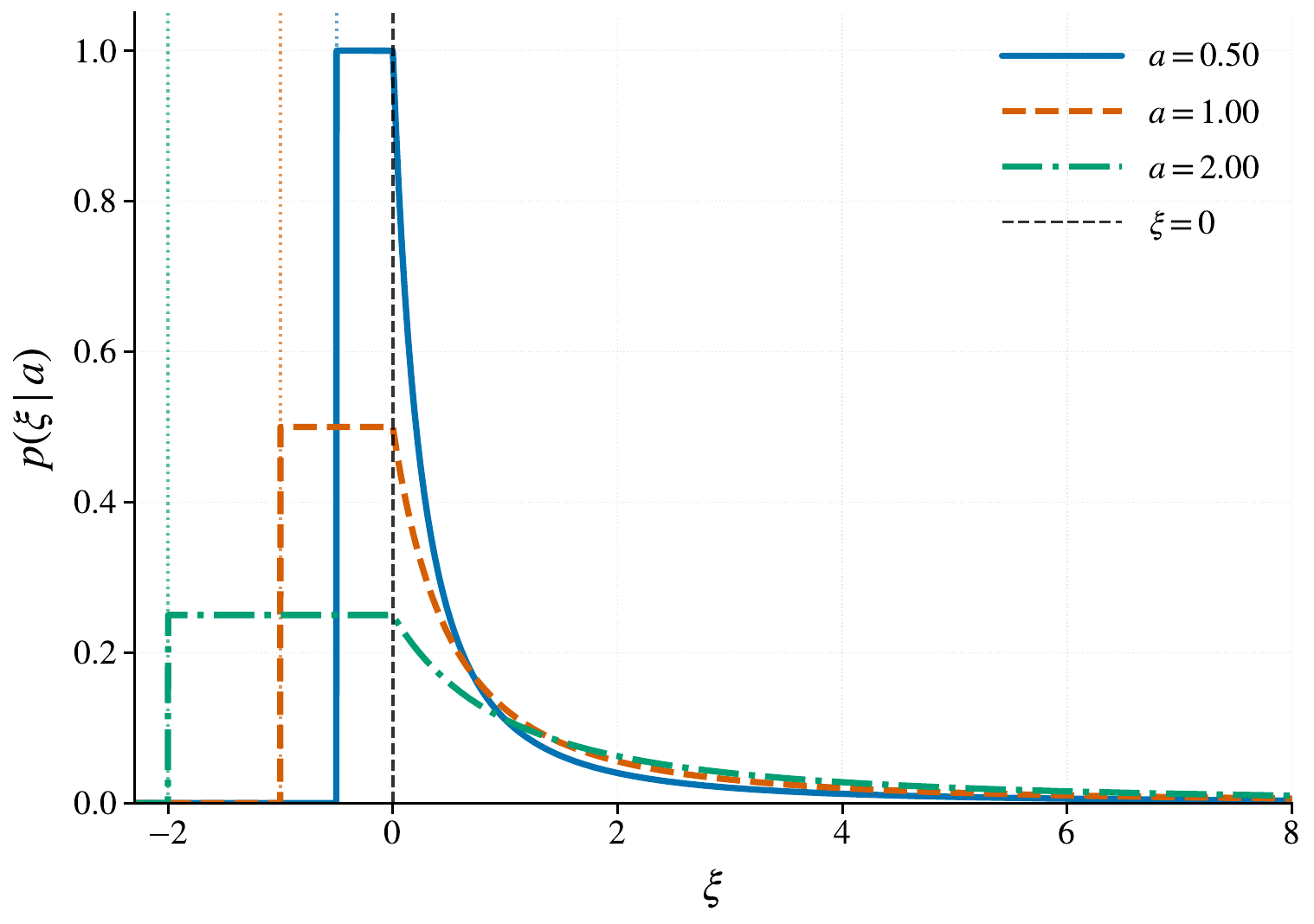}
    \caption{
        Induced prior density of the extreme value index $\xi$
        under the transformation $\xi=a(e^\zeta-1)$, with
        $\zeta\sim\operatorname{Laplace}(0,1)$.
        The three curves correspond to $a=0.50$ (blue), $a=1.00$
        (orange), and $a=2.00$ (green). The vertical dashed line
        at $\xi=0$ denotes the exponential-tail case, while the
        dotted vertical lines indicate the lower support boundaries,
        $\xi=-a$.
    }
    \label{fig:induced_prior_xi}
\end{figure}

The transformation accommodates the three relevant tail regimes:
\begin{equation}
\begin{aligned}
\xi<0 &\quad \Longrightarrow \quad \text{bounded upper endpoint},\\
\xi=0 &\quad \Longrightarrow \quad \text{exponential tail},\\
\xi>0 &\quad \Longrightarrow \quad \text{heavy tail}.
\end{aligned}
\end{equation}
Moreover, the centered Laplace prior on $\zeta$ assigns equal probability to positive and negative values. Since the transformation is strictly increasing and maps zero to zero, the induced prior satisfies
\begin{equation}
\Pr(\xi<0\mid\sigma)
=
\Pr(\xi>0\mid\sigma)
=
\frac{1}{2}.
\end{equation}
Hence, the prior allocates equal probability to the bounded-tail and heavy-tailed regimes, while placing their boundary at the exponential-tail case. This balanced allocation in probability holds despite the asymmetric support and shape of the induced density, as illustrated in Figure~\ref{fig:induced_prior_xi}.

Posterior inference is performed using Hamiltonian Monte Carlo (HMC)~\cite{Brooks_2011}, as implemented in \texttt{Stan}~\cite{stanDevelopmentTeam2025}. Let
\begin{equation}
\mathbf{L}=(L_1,\ldots,L_M)
\end{equation}
denote the vector of latent representative locations. By incorporating these locations explicitly into the hierarchical model, posterior uncertainty associated with the within-interval evaluation points is propagated to the estimates of $\sigma$, $\xi$, and $\kappa$. This is particularly important when the partition intervals are sufficiently wide that replacing each interval with a fixed representative point could introduce appreciable discretization bias.

Posterior samples provide estimates of the model parameters and, through the EGP hazard, the corresponding hazard and survival functions, together with their posterior uncertainty. In particular, inference on tail behaviour is obtained directly from the posterior distribution of $\xi$, allowing uncertainty about the bounded-tail, exponential-tail, and heavy-tailed regimes to be quantified.

\section{Model with tail covariates}
\label{sec:LookAtMyTail}

A key advantage of the proposed framework is its flexibility in allowing
covariate effects to act on both the scale of the hazard and the tail
behaviour governing extreme survival times. We extend the baseline model
by allowing the scale parameter $\sigma$ and the extreme value index
$\xi$ to depend on covariates. Thus, both the overall time scale of the
survival distribution and its asymptotic tail behaviour may vary across
individuals with different covariate profiles.

Let $\boldsymbol{Z}=(Z_1,\ldots,Z_p)^\top$ denote a vector of $p$
covariates. The scale parameter is modelled using a log-linear
specification,
\begin{equation}
\label{eq:sigma_covariates}
\sigma(\boldsymbol{Z})
=
\exp\left(
\alpha_0+\sum_{k=1}^{q}\alpha_k Z_k
\right),
\qquad q\leq p,
\end{equation}
which guarantees a positive scale for every covariate profile. The
coefficients $\alpha_0,\ldots,\alpha_q$ describe the multiplicative
effects of the covariates on the overall time scale of the EGP
distribution.

The extreme value index is also allowed to depend on the covariates.
However, its parameterization must respect the support constraint of
the GP distribution. Specifically, the condition
\begin{equation}
\label{eq:xi_support_covariates}
1+\frac{\xi(\boldsymbol{Z})}{\sigma(\boldsymbol{Z})}t>0,
\qquad t\in\mathcal{T},
\end{equation}
requires
\begin{equation}
\label{eq:xi_lower_covariates}
\xi(\boldsymbol{Z})
>
-\frac{\sigma(\boldsymbol{Z})}{\sup\mathcal{T}},
\end{equation}
assuming that $\mathcal{T}$ has a finite, positive supremum and
contains zero. This constraint must hold for every covariate profile
in the relevant covariate space.

To construct a valid parameterization, assume that the covariates
are normalized to $[0,1]$. Let $\mathcal{C}\subseteq[0,1]^p$ denote
the relevant covariate space and define the common lower-bound
parameter
\begin{equation}
\label{eq:common_lower_bound}
a_\sigma
=
\inf_{\boldsymbol{z}\in\mathcal{C}}
\frac{\sigma(\boldsymbol{z})}{\sup\mathcal{T}}.
\end{equation}
Provided that $a_\sigma>0$, this bound ensures that satisfying
$\xi(\boldsymbol{Z})>-a_\sigma$ is sufficient to guarantee
the support condition in Equation~\eqref{eq:xi_lower_covariates}
for every covariate profile.

Under the log-linear scale specification in
Equation~\eqref{eq:sigma_covariates}, when the first $q$ covariates
range independently over $[0,1]$, the lower bound can be written as
\begin{equation}
\label{eq:common_lower_bound_explicit}
a_\sigma
=
\frac{1}{\sup\mathcal{T}}
\exp\left(
\alpha_0+\sum_{k=1}^{q}\min(0,\alpha_k)
\right).
\end{equation}

We introduce a sequential parameterization of the extreme value index
using latent variables
\begin{equation}
\label{eq:zeta_prior_covariates}
\zeta_k\overset{\mathrm{ind}}{\sim}
\operatorname{Laplace}(0,1),
\qquad k=0,\ldots,p.
\end{equation}
First, define the initial value
\begin{equation}
\label{eq:tau0_covariates}
\tau_0
=
a_\sigma\left(e^{\zeta_0}-1\right).
\end{equation}
This transformation guarantees that $\tau_0>-a_\sigma$ for every
finite $\zeta_0$.

Next, for $k=1,\ldots,p$, define the recursive sequence
\begin{equation}
\label{eq:tau_recursion_covariates}
\tau_k(\boldsymbol{Z})
=
\tau_{k-1}(\boldsymbol{Z})
+
\left\{
-\tau_{k-1}(\boldsymbol{Z})-a_\sigma
\right\}
\left(1-e^{\zeta_k}\right)Z_k,
\end{equation}
where $\tau_0(\boldsymbol{Z})=\tau_0$. Equivalently, the recursion
can be expressed as
\begin{equation}
\label{eq:tau_recursion_factorized}
\tau_k(\boldsymbol{Z})+a_\sigma
=
\left\{\tau_{k-1}(\boldsymbol{Z})+a_\sigma\right\}
\left\{1-Z_k+Z_k e^{\zeta_k}\right\}.
\end{equation}
Since $Z_k\in[0,1]$ and $e^{\zeta_k}>0$, the multiplicative factor
in Equation~\eqref{eq:tau_recursion_factorized} is strictly positive.
It follows recursively that
\begin{equation}
\label{eq:tau_support_covariates}
\tau_k(\boldsymbol{Z})>-a_\sigma,
\qquad k=0,\ldots,p.
\end{equation}

The final covariate-dependent extreme value index is therefore
defined as
\begin{equation}
\label{eq:xi_final_covariates}
\xi(\boldsymbol{Z})=\tau_p(\boldsymbol{Z}).
\end{equation}
By construction,
\begin{equation}
\xi(\boldsymbol{Z})
>
-a_\sigma
\geq
-\frac{\sigma(\boldsymbol{Z})}{\sup\mathcal{T}},
\end{equation}
so the GP support condition is satisfied throughout the specified
covariate space.

This construction allows the extreme value index to take negative,
zero, or positive values, corresponding respectively to bounded-tail,
exponential-tail, and heavy-tailed behaviour. The sequential
parameterization preserves the support constraint without restricting
the model to a single tail regime in advance.

Combining the covariate-dependent scale and extreme value index with
the latent representative locations introduced in
Section~\ref{sec:MYMODEL}, the second-stage model becomes
\begin{equation}
\label{eq:egp_covariate_likelihood}
\widehat h_j
\mid
L_j,\boldsymbol{Z}_j,
\boldsymbol{\alpha},\boldsymbol{\zeta},\kappa
\sim
\mathcal{N}^{+}
\left(
h_{\mathrm{EGP}}\left(
L_j;
\sigma(\boldsymbol{Z}_j),
\xi(\boldsymbol{Z}_j),
\kappa
\right),
\widehat{\tau}_j^2
\right),
\end{equation}
where $\sigma(\boldsymbol{Z}_j)$ is given by
Equation~\eqref{eq:sigma_covariates}, $\xi(\boldsymbol{Z}_j)$ is
defined recursively through Equations~\eqref{eq:tau0_covariates}
and~\eqref{eq:tau_recursion_covariates}, and
$\boldsymbol{\zeta}=(\zeta_0,\ldots,\zeta_p)^\top$ denotes the
vector of latent variables.

The resulting model allows covariates to affect both the overall
scale of survival times and their asymptotic tail behaviour, while
accounting for uncertainty in the latent representative locations
and propagating first-stage estimation uncertainty through
$\widehat{\tau}_j^2$.

\section{Bias correction}
\label{sec:bias_correction}

The piecewise exponential approximation introduces discretization bias when
the underlying hazard varies appreciably within an interval. In the
proposed framework, two related mechanisms are of particular relevance.
The first arises from representing a continuously varying hazard by a
single constant value within each interval. The second is associated with
the highly heterogeneous interval widths and exposure times that can arise
in heavy-tailed settings, where a small number of extreme observations may
strongly influence the final intervals of the partition.

We considered three strategies to mitigate these effects. The first is a best representative approach, in which each interval is represented by a deterministic location chosen so that the hazard at that location matches the average hazard over the interval. This approach was previously implemented in Janotti et al.~\cite{janotti2026beyond}. Although it is theoretically appealing as a correction for within-interval curvature, it did not exhibit the intended behavior in the resulting model and was therefore abandoned. The second strategy is an adaptive partition refinement, in which the interval boundaries are determined using a preliminary estimate of the underlying distribution. This strategy was also considered in Janotti et al.~\cite{janotti2026beyond} and is retained here, with an important simplification: rather than repeatedly updating the partition as part of an iterative estimation scheme, we use only a first-step approximation based on an initial exponential fit. This avoids the additional computational cost and feedback associated with iterative partition updating. The third strategy, and the main approach considered in this work, is a latent location formulation. Instead of deterministically selecting a single representative point within each interval, the representative location is treated as an unobserved quantity and its uncertainty is incorporated directly into the hierarchical model. Before introducing this formulation, we describe the two sources of bias and the two alternative strategies that motivate it.

\subsection{Sources of discretization bias}

Consider an interval
\begin{equation}
I_j=(\ell_j,u_j),
\qquad
\ell_j=\inf I_j,
\qquad
u_j=\sup I_j.
\end{equation}

If the true hazard is $h(t)$, the expected number of events per unit
exposure over the interval is determined by its average hazard,
\begin{equation}
\label{eq:average_hazard}
c_j
=
\frac{1}{u_j-\ell_j}
\int_{\ell_j}^{u_j} h(t)\,\mathrm{d}t.
\end{equation}

For the GP hazard,
\begin{equation}
\label{eq:gp_hazard_bias}
h(t)
=
\frac{1}{\sigma+\xi t},
\qquad
1+\frac{\xi t}{\sigma}>0,
\end{equation}
the corresponding average hazard is
\begin{equation}
\label{eq:gp_average_hazard}
c_{j,\xi}
=
\frac{1}{u_j-\ell_j}
\frac{1}{\xi}
\log\left(
\frac{\sigma+\xi u_j}
{\sigma+\xi\ell_j}
\right),
\qquad
\xi\neq0.
\end{equation}
For $\xi=0$, the expression is obtained by continuity and reduces to
\begin{equation}
c_{j,0}=\frac{1}{\sigma}.
\end{equation}
The derivation of this expression is provided in Appendix~B.

In the piecewise exponential approximation, the continuously varying
hazard within $I_j$ is represented by a single value. A conventional
choice is to evaluate the hazard at the midpoint,
\begin{equation}
m_j=\frac{\ell_j+u_j}{2},
\end{equation}
and use $h(m_j)$ as the representative hazard. However, the
interval-level hazard estimated from the piecewise exponential model
corresponds to the average hazard $c_{j,\xi}$, which generally differs
from the midpoint value:
\begin{equation}
h(m_j)\neq c_{j,\xi}.
\end{equation}

This discrepancy is a discretization effect arising from the variation
of the hazard within the interval. A natural deterministic correction
is therefore to select a representative location
$r_{j,\xi}\in I_j$ satisfying
\begin{equation}
\label{eq:representative_hazard_matching}
h(r_{j,\xi})=c_{j,\xi}.
\end{equation}
For the GP hazard, solving this equation gives
\begin{equation}
\label{eq:representative_location_gp}
r_{j,\xi}
=
\frac{1}{\xi}
\left(
\frac{1}{c_{j,\xi}}-\sigma
\right),
\qquad
\xi\neq0.
\end{equation}

When $\xi=0$, the hazard is constant throughout the interval, so the
choice of representative location is immaterial. The midpoint may
therefore be used:
\begin{equation}
r_{j,0}=m_j=\frac{\ell_j+u_j}{2}.
\end{equation}

The discrepancy between midpoint evaluation and the representative-point
construction can be characterized analytically for an interval of width
$L$ beginning at zero, $I_1=[0,L)$. For the GP hazard, the representative
point $r_1$, defined by matching the hazard at that point to the
interval-average hazard, is
\begin{equation}
\label{eq:representative_point}
r_1
=
\frac{L}{\log\left(1+\frac{\xi L}{\sigma}\right)}
-\frac{\sigma}{\xi},
\qquad
m_1=\frac{L}{2}.
\end{equation}

Introducing the dimensionless quantity
\begin{equation}
\label{eq:dimensionless_width}
z=\frac{\xi L}{\sigma},
\end{equation}
the relative location of the representative point becomes
\begin{equation}
\label{eq:relative_location}
R(z)
=
\frac{r_1}{m_1}
=
2\left(
\frac{1}{\log(1+z)}-\frac{1}{z}
\right).
\end{equation}
As derived in Appendix~B, its expansion around zero is
\begin{equation}
\label{eq:R_expansion}
R(z)=1-\frac{z}{6}+O(z^2),
\qquad z\to 0.
\end{equation}
Thus, the representative point approaches the midpoint as the interval
width becomes small relative to the scale of hazard variation.
Furthermore, $R(z)<1$ for $z>0$ and $R(z)>1$ for $-1<z<0$.
Consequently, positive values of $\xi$ place the representative point
to the left of the midpoint, whereas negative values place it to the
right.

This discrepancy may affect the estimation of tail behaviour. Let
$\xi'$ denote the index obtained by matching the midpoint hazard to
the interval-average hazard. Under the corresponding hazard-matching
relationship,
\begin{equation}
\label{eq:hazard_matching}
h(r_1;\sigma,\xi)=h(m_1;\sigma,\xi'),
\end{equation}
the resulting index is claimed to satisfy
\begin{equation}
\label{eq:xi_prime}
\xi'=\xi R(z).
\end{equation}
Under this relationship, the relative location of the representative
point implies
\begin{equation}
\label{eq:xi_bias}
\begin{cases}
\xi'<\xi, & \xi>0,\\
\xi'>\xi, & \xi<0.
\end{cases}
\end{equation}
Hence, midpoint representation shifts the matched index towards zero,
understating positive tail heaviness and making negative indices less
negative. The magnitude of the discrepancy depends on the relative
variation of the hazard within the interval, as characterized by
$|\xi|L/\sigma$.

The representative-point construction addresses this discrepancy at
the interval level by defining $r_j$ such that
\begin{equation}
\label{eq:representative_matching}
h(r_j;\sigma,\xi)=c_j,
\end{equation}
where $c_j$ denotes the interval-average hazard. This provides a direct
correspondence between the continuous GP hazard and the piecewise
exponential approximation. However, because $r_j$ depends
deterministically on the unknown parameters $(\sigma,\xi)$,
incorporating this correction introduces additional dependence into
the estimation procedure. Furthermore, its empirical performance was
unsatisfactory in the previous implementation. Accordingly, the
deterministic representative-point correction was considered but
ultimately discarded as a correction mechanism in
Janotti et al.~\cite{janotti2026beyond}. The present approach instead
modifies the partition and subsequently treats the representative
locations probabilistically.

\begin{figure}[htbp]
    \centering
    \includegraphics[
        width=0.95\textwidth
    ]{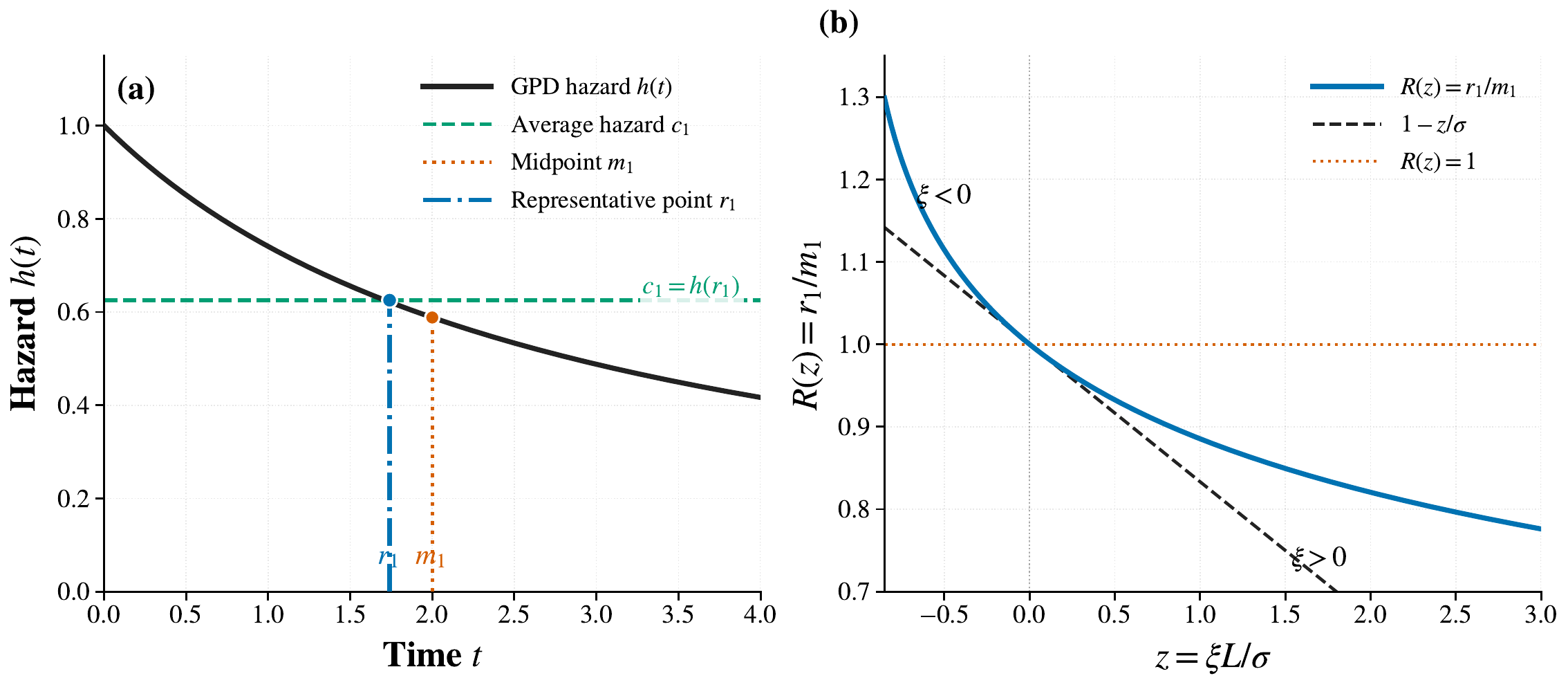}
    \caption{
        Illustration of the discretization effect induced by midpoint
        representation of the GP hazard. Panel~(a) shows the GP hazard
        $h(t)=1/(\sigma+\xi t)$ over the interval $I_1=[0,L)$ for
        $\sigma=1$ and $\xi>0$, together with the interval-average hazard
        $c_1$. The midpoint $m_1=L/2$ generally differs from the
        representative point $r_1$ satisfying $h(r_1)=c_1$. Panel~(b)
        shows the ratio $R(z)=r_1/m_1$, where
        $z=\xi L/\sigma$. The ratio satisfies $R(z)<1$ for $\xi>0$ and
        $R(z)>1$ for $\xi<0$, while $R(z)\rightarrow1$ as $z\rightarrow0$.
        Thus, midpoint evaluation systematically shifts the representative
        location away from the point corresponding to the interval-average
        hazard.
    }
    \label{fig:GP_discretization_bias}
\end{figure}

Figure~\ref{fig:GP_discretization_bias} illustrates the discrepancy
between midpoint and representative-point evaluation and its
dependence on the dimensionless interval width $z$. For positive
$\xi$, the decreasing GP hazard places the representative point to
the left of the midpoint, with the displacement diminishing as
$z$ approaches zero.

\subsubsection{Tail amplification}

A second source of instability becomes important in heavy-tailed regimes.
For a $\operatorname{GP}(\sigma,\xi)$ random variable with $\xi>0$, the
$k$th central moment exists if and only if $
\xi<1/k.$ This property has direct implications for finite samples. Let
$
X_1,\ldots,X_n\sim\operatorname{GP}(\sigma,\xi)
$
and denote the associated order statistics by
$
X_{(1)}\leq\cdots\leq X_{(n)}.
$ If $\xi\geq1$, then
$
\mathbb{E}[X_{(n)}]=\infty.
$  More generally, sufficiently high-order moments of the largest order
statistics do not exist. Consequently, extreme observations can have a
disproportionate influence on the range of the observed data.

This phenomenon affects the construction of the final intervals in a
piecewise exponential representation. If partition boundaries are
determined directly from the observed extremes, a small number of very
large observations can generate intervals with unusually large widths.
The corresponding exposure times can then become very large relative to
the number of observed events. As introduced in section \ref{subsec:2-stage}, the piecewise exponential estimator in interval $I_j$ is $ \widehat h_j =D_j/E_j$,
 where $D_j$ is the number of observed events in the interval and $E_j$ is the corresponding total exposure time. If the final interval contains only a small number of events but a very large amount of exposure, then $\widehat h_j$ can become artificially small. Thus, heavy-tailed observations can amplify the discretization effects already present in the piecewise exponential approximation.

The two mechanisms are therefore complementary. Curvature bias results
from representing a varying hazard by a single evaluation point, whereas
tail amplification results from extreme observations producing highly
unequal interval widths and exposure times. The first motivates correcting
the representative location, while the second motivates adapting the
partition.

\subsection{Adaptive partition refinement}

A complementary strategy is to modify the partition rather than
the representative location. Instead of fixing the interval boundaries
independently of the data, they can be chosen according to a preliminary
estimate of the underlying distribution. The objective is to distribute
the observations more evenly across the intervals and thereby reduce
the occurrence of extremely heterogeneous exposure times.

In principle, the partition can be refined using the fitted GP
CDF
$F_{\mathrm{GP}}(\cdot;\widehat{\sigma},\widehat{\xi})$.
For $M$ intervals, the boundaries can be chosen approximately
according to GP quantiles:
\begin{equation}
\label{eq:gp_adaptive_boundaries}
\rho_j
=
F_{\mathrm{GP}}^{-1}
\left(
\frac{j}{M};
\widehat{\sigma},\widehat{\xi}
\right),
\qquad
j=0,\ldots,M,
\end{equation}
subject to the relevant truncation and time-domain constraints.
This construction assigns approximately equal probability mass to
each interval:
\begin{equation}
\label{eq:gp_equal_probability}
F_{\mathrm{GP}}(\rho_{j+1};\widehat{\sigma},\widehat{\xi})
-
F_{\mathrm{GP}}(\rho_j;\widehat{\sigma},\widehat{\xi})
\approx
\frac{1}{M},
\qquad
j=0,\ldots,M-1.
\end{equation}

An adaptive construction of this type was considered and implemented
in Janotti et al.~\cite{janotti2026beyond}.

In the present work, however, we do not use a fully iterative
refinement scheme. Instead, we use a first-step approximation in
which the event times are initially assumed to follow an exponential
distribution,
\(
X\sim\operatorname{Exp}(\lambda),
\)
with right censoring. The rate parameter is estimated by maximum
likelihood from the available censored observations. In particular,
if $D$ denotes the total number of observed events and $E$ denotes
the total exposure time, the maximum likelihood estimator is
\begin{equation}
\label{eq:initial_exponential_rate}
\widehat{\lambda}_0=\frac{D}{E}.
\end{equation}
This estimate is then used to construct the partition before fitting
the final model. Thus, the partition is determined once, using
$\widehat{\lambda}_0$, and is not updated during estimation of the
final model.

The corresponding estimated exponential CDF is
\begin{equation}
\label{eq:initial_exponential_cdf}
\widehat{F}_0(t)
=
1-\exp(-\widehat{\lambda}_0t),
\qquad t\geq 0.
\end{equation}
The associated quantile function is
\begin{equation}
\label{eq:initial_exponential_quantiles}
\rho_j
=
\widehat{F}_0^{-1}\left(\frac{j}{M}\right)
=
-\frac{1}{\widehat{\lambda}_0}
\log\left(1-\frac{j}{M}\right),
\qquad j=0,\ldots,M-1.
\end{equation}
Because the untruncated exponential distribution has an unbounded
support, the quantile at $j=M$ is infinite. In practice, the final
boundary must therefore be set by a finite upper time limit, or the
quantile construction must be applied to a suitably truncated
exponential distribution. The resulting boundaries must also respect
the time-domain constraints of the analysis.

This one-step construction provides a simple data-adaptive partition
without introducing an additional iterative estimation loop. The
partition is determined independently of the parameters estimated
by the final model, avoiding the need to repeatedly reconstruct the
exposure terms and refit the model as the parameter estimates change.
The approach therefore retains the principal benefit of adaptive
partitioning while keeping the computational structure of the
proposed model fixed.

The adaptive partition addresses primarily the second source of
discretization instability, namely the amplification caused by highly
unequal interval widths in the presence of extreme observations. It
does not, however, explicitly account for uncertainty about the
appropriate evaluation point within each interval. This motivates
the third strategy, which is the main bias-correction approach
adopted in this work.

\subsection{Latent location correction}
\label{sec:latent_location}

Rather than deterministically replacing the midpoint by a bias-corrected
representative, we treat the representative location within each interval
as a latent quantity. This allows uncertainty about the appropriate
evaluation point to be propagated through the hierarchical model without
requiring an iterative correction procedure.

The deterministic representative point introduced above provides the
theoretical motivation for this construction. In particular, for the GP
hazard, the point $r_{j, \xi}$ satisfying $
h(r_{j, \xi})=c_{j, \xi}$, identifies the location at which the continuous hazard coincides with the
average hazard represented by the piecewise exponential estimate. However,
using $r_{j, \xi}$ directly would require its value to be updated as the unknown
model parameters change.

Instead, we use a simple probabilistic specification for the latent
representative locations. For each interval $I_j=(\ell_j,u_j)$, let
\begin{equation}
L_j=\ell_j+B_j(u_j-\ell_j),
\qquad
B_j\sim\operatorname{Uniform}(0,1).
\end{equation}
Equivalently,
\begin{equation}
L_j\sim\operatorname{Uniform}(\ell_j,u_j).
\end{equation}
Thus, all locations within an interval are initially assigned equal
prior density.

Unlike fixed midpoint evaluation, this specification allows the
representative locations to vary within their intervals. The data
inform their posterior distributions, and uncertainty in
$L_1,\ldots,L_M$ is propagated into inference on
$(\sigma,\xi,\kappa)$. This provides a flexible alternative to fixing
each interval-level hazard estimate $\widehat{h}_j$ at the midpoint
$(\ell_j+u_j)/2$, although it does not guarantee the elimination of
discretization bias.

A more informative alternative would center the location distribution
around a deterministic representative point $r_j$. For example, the
standardized location
\begin{equation}
B_j=\frac{L_j-\ell_j}{u_j-\ell_j}
\end{equation}
could follow a beta distribution with parameters chosen to concentrate
its mass around the standardized value
$(r_j-\ell_j)/(u_j-\ell_j)$. Such a specification would allow the
degree of concentration around $r_j$ to be controlled, but would
introduce additional modeling choices. We therefore retain the uniform
specification as a simple baseline. More elaborate adaptive or
iterative approaches to modeling the latent locations are left for
future work.

\section{Simulation}
\label{sec:simulation}

We conduct two simulation studies to evaluate the finite-sample performance
and robustness of the proposed estimation procedure. The first study
investigates the effect of the number of bins on posterior uncertainty
quantification under correctly specified GP models, while varying the
sample size, censoring proportion, and tail behaviour of the censoring
distribution. The second study evaluates robustness to model
misspecification by generating survival and censoring times from a
lognormal distribution.

\subsection{Simulation study I: choice of the number of bins}

The first simulation study evaluates the effect of the number of bins on
the inferential performance of the proposed piecewise exponential
approximation. In particular, we investigate whether a relatively small
number of bins is sufficient to recover the true GP parameters and provide
adequate posterior coverage.

For each simulation replicate $i$, let
\(
\left\{(X_{i,j},C_{i,j}):j=1,\ldots,n\right\}
\)
denote a sample of independent survival and censoring times. The survival
times $\{X_{i,j}\}_{j=1}^n$ are independent and identically distributed,
as are the censoring times $\{C_{i,j}\}_{j=1}^n$, and the two sequences are
mutually independent. Specifically,
\(
X_{i,j}\sim\operatorname{GP}(1,\xi),
\) \(
C_{i,j}\sim\operatorname{GP}(\sigma^c,\xi^c),
\)
where $\xi$ and $\xi^c$ are the shape parameters of the survival and
censoring distributions, respectively, and $\sigma^c$ is the scale
parameter of the censoring distribution.

The observed follow-up time and censoring indicator are defined as
\(
T_{i,j}
=
\min(X_{i,j},C_{i,j}),
\) \(
\delta_{i,j}
=
\mathds{1}\{X_{i,j}\leq C_{i,j}\}.
\)

Thus, $\delta_{i,j}=1$ indicates an observed event, whereas
$\delta_{i,j}=0$ indicates right censoring.

We consider the shape parameters
\(
\xi\in\{-2,0,2\},
\) \(
\xi^c\in\{-2,0,2\}.
\)
The censoring scale parameter $\sigma^c$ is selected separately for each
combination of $(\xi,\xi^c)$ to produce approximately $25\%$, $50\%$, and
$75\%$ censoring. In addition to varying the overall censoring proportion, we vary the tail
behaviour of the censoring distribution through $\xi^c$. These two aspects
of censoring are not equivalent: two censoring mechanisms can produce
similar overall proportions of censored observations while generating very
different patterns of censoring over time. The values used in the simulation are reported in
Table~\ref{tab:scaleC}.

\begin{table}[ht]
\centering
\caption{Values of the censoring scale parameter $\sigma^c$ used to obtain
approximately $25\%$, $50\%$, and $75\%$ censoring for each combination of
the survival and censoring shape parameters $(\xi,\xi^c)$.}
\label{tab:scaleC}
\begin{tabular}{ccccc}
\toprule
\multirow{2}{*}{$\xi$} &
\multirow{2}{*}{$\xi^c$} &
\multicolumn{3}{c}{$\sigma^c$} \\
\cmidrule(lr){3-5}
& & 25\% & 50\% & 75\% \\
\midrule
-2 & -2 & 1.6 & 1.0 & 0.6 \\
0  & -2 & 5.2 & 2.3 & 0.9 \\
2  & -2 & 31.0 & 5.4 & 1.3 \\
-2 &  0 & 1.1 & 0.4 & 0.2 \\
0  &  0 & 3.0 & 1.0 & 0.3 \\
2  &  0 & 17.3 & 2.6 & 0.5 \\
-2 &  2 & 0.8 & 0.2 & 0.03 \\
0  &  2 & 1.9 & 0.4 & 0.06 \\
2  &  2 & 9.1 & 1.0 & 0.11 \\
\bottomrule
\end{tabular}
\end{table}

The choice of $\xi^c$ determines the tail behaviour of the censoring
mechanism and therefore the extent to which subjects remain under
observation for long periods. When $\xi^c<0$, the censoring distribution
has a finite upper endpoint, implying that follow-up cannot extend beyond a
finite time horizon. This setting is representative of administrative
censoring or fixed study-end designs. In particular, when the upper
endpoint is common across subjects, it is closely related to the classical
Type~I (administrative) censoring scheme, in which all individuals are
censored at a predetermined study-ending time. More generally, the
negative-shape GP allows the censoring times to vary while still imposing a
finite maximum follow-up time. If $\xi^c=0$, the censoring times follow an exponential distribution and
the censoring hazard is constant over time. This provides a memoryless
censoring mechanism in which the instantaneous chance of being censored
does not depend on the duration of follow-up. Unlike the finite-endpoint
case, there is no predetermined maximum censoring time, although long
follow-up times become progressively less likely. Lastly,  when $\xi^c>0$, the censoring distribution is heavy-tailed and has no finite
upper endpoint. Consequently, some individuals may remain under
observation for very long periods, increasing the probability that extreme
survival times are observed rather than censored. This setting represents a
more favourable observation regime for tail inference, because the
censoring mechanism itself does not impose a finite limit on follow-up and
has a heavier tail than the exponential case.
These three scenarios therefore represent progressively different
follow-up regimes: finite-horizon administrative censoring for
$\xi^c<0$, memoryless censoring for $\xi^c=0$, and heavy-tailed,
potentially very long follow-up for $\xi^c>0$. Comparing them allows the
proposed method to be evaluated under qualitatively distinct censoring
mechanisms and, importantly, under different degrees of difficulty in
recovering the extreme tail of the survival distribution.

We consider two sample sizes,
\(
n\in\{100,1000\},
\)
and, for the proposed model, four possible numbers of bins,
\(
M\in\{5,10,15,20\}.
\)
For this simulation study, we fix $\kappa=1,$so that the EGP reduces to the classical GP. This choice isolates the effect of the piecewise exponential approximation and allows the simulation to focus on the estimation of the scale parameter $\sigma$ and the extreme value index $\xi$ under different censoring configurations. The effect of estimating $\kappa$ is investigated separately in the second simulation study.

\begin{figure}[ht]
\centering
\includegraphics[width=0.99\linewidth]{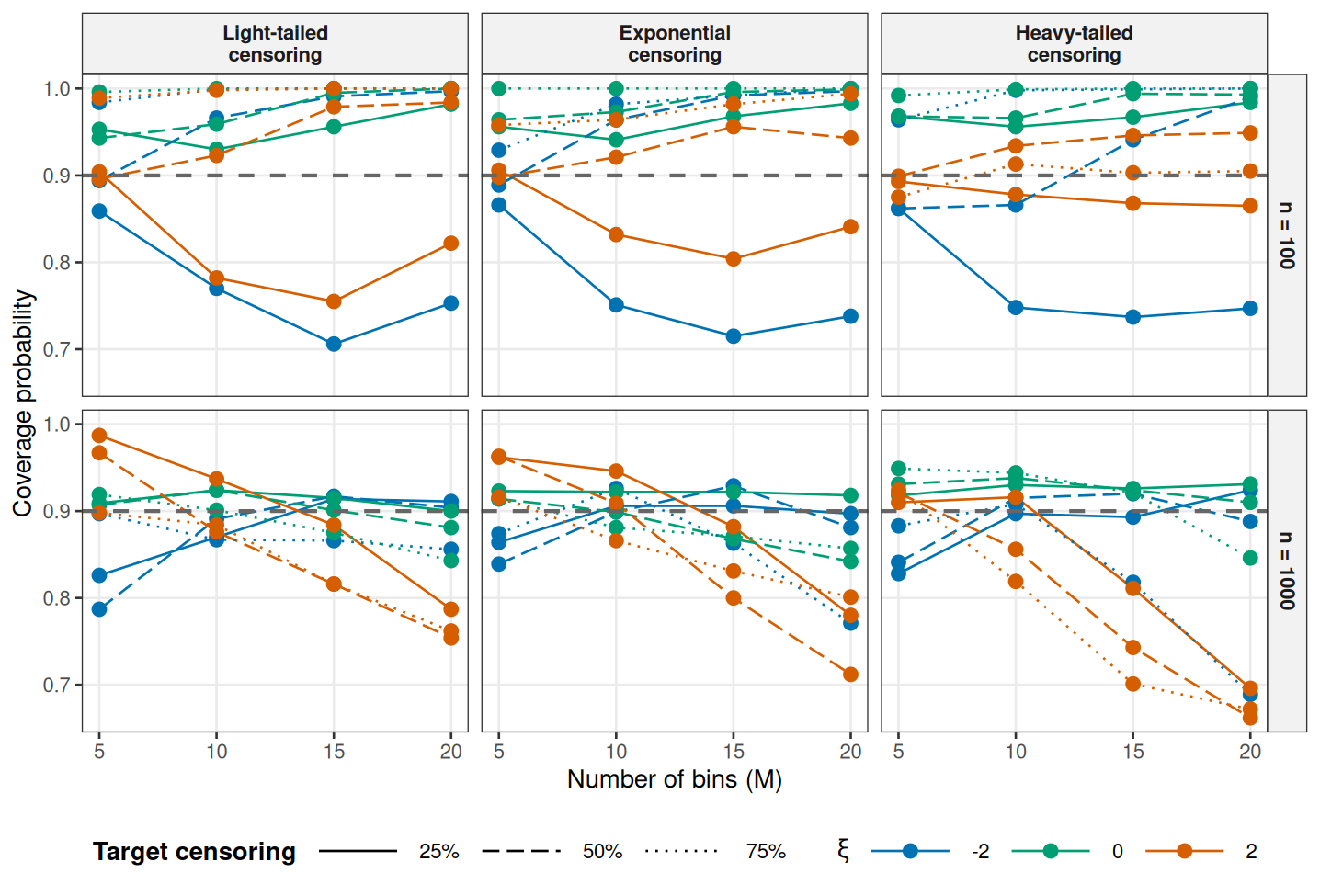}
\caption{Empirical coverage probability of the $90\%$ posterior credible
interval for $\xi$ as a function of the number of bins. Panels are
stratified by sample size and the tail behaviour of the censoring
distribution, while colours and line types represent the true value of
$\xi$ and the target censoring proportion, respectively. The dashed
horizontal line indicates the nominal coverage level of $90\%$.}
\label{fig:coverage_xi}
\end{figure}

\begin{figure}[ht]
\centering
\includegraphics[width=0.99\linewidth]{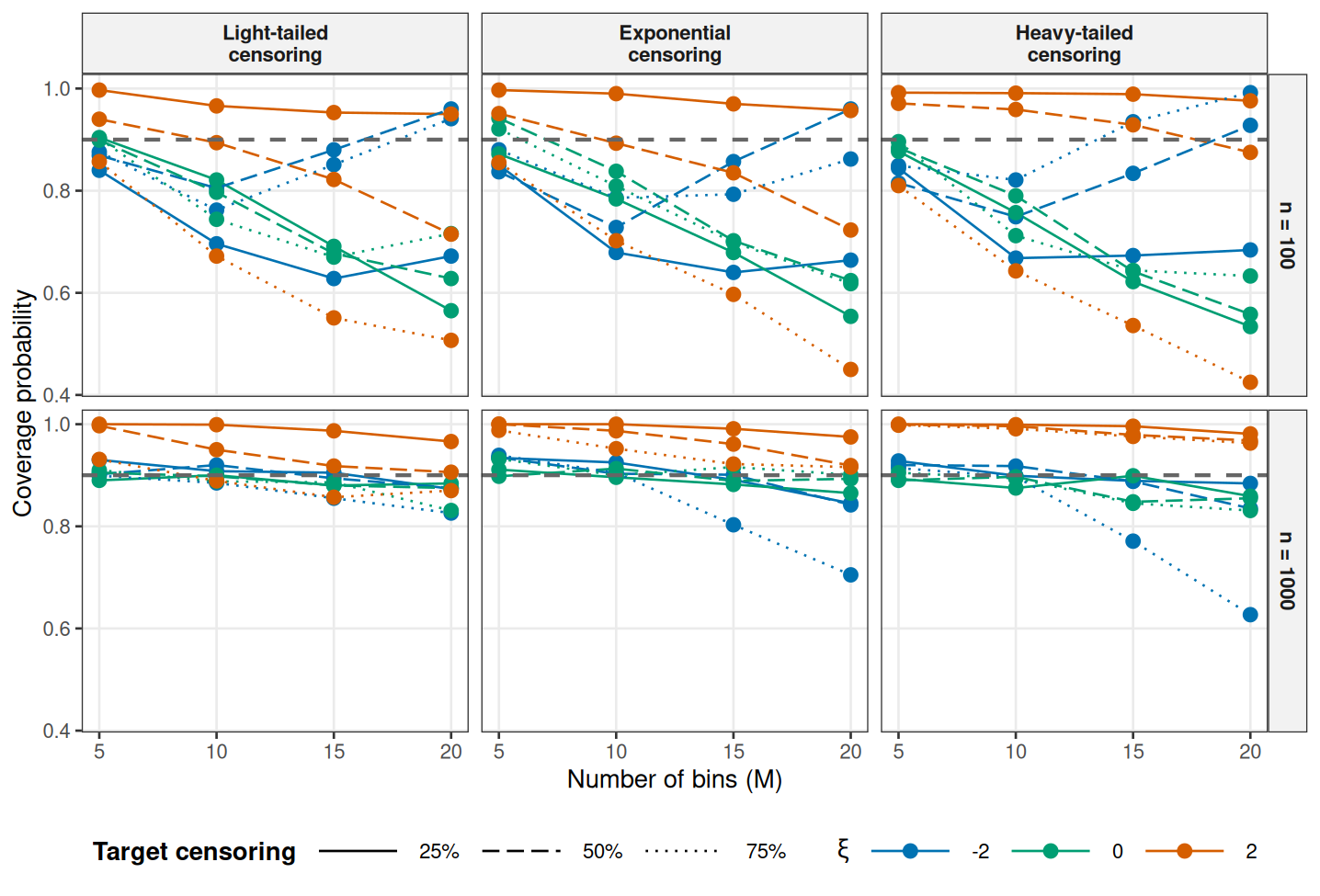}
\caption{Empirical coverage probability of the $90\%$ posterior credible
interval for $\sigma$ as a function of the number of bins. Panels are
stratified by sample size and the tail behaviour of the censoring
distribution, while colours and line types represent the true value of
$\xi$ and the target censoring proportion, respectively. The dashed
horizontal line indicates the nominal coverage level of $90\%$.}
\label{fig:coverage_sigma}
\end{figure}

Figures~\ref{fig:coverage_xi} and~\ref{fig:coverage_sigma} summarize the
empirical coverage probabilities of the $90\%$ posterior credible intervals
for $\xi$ and $\sigma$, respectively.

The simulation results indicate that a relatively small number of bins
provides reliable inferential performance across the scenarios considered.
In particular, $M=5$ and $M=10$ generally produce empirical coverage
probabilities close to the nominal $90\%$ level for both parameters and
across the different sample sizes, censoring proportions, and censoring
mechanisms. Although coverage tends to become more variable as the number
of bins increases, the deterioration is gradual over the range considered.

For the shape parameter $\xi$ (Figure~\ref{fig:coverage_xi}), the effect of
increasing the number of bins depends on both the sample size and the
censoring mechanism. For $n=100$ under bounded censoring
($\xi^c=-2$), increasing $M$ generally improves empirical coverage, except
in scenarios with relatively low censoring and extreme values of $\xi$,
namely $\xi=-2$ and $\xi=2$. Under exponential censoring
($\xi^c=0$), a similar qualitative pattern is observed, although the
improvement is less pronounced. Under heavy-tailed censoring
($\xi^c=2$), coverage remains comparatively stable as $M$ increases, with
the main exception being the case $\xi=-2$ under $25\%$ censoring, where a
decrease in coverage is observed.

For $n=1000$, the empirical coverage of $\xi$ is generally more stable
across the different choices of $M$. Under both bounded and exponential
censoring, changing the number of bins has a limited effect on coverage,
with a slight deterioration for $\xi=2$ and a small improvement for
$\xi=-2$. Under heavy-tailed censoring, the same qualitative pattern
persists, although the reduction in coverage for $\xi=2$ becomes more
pronounced as the number of bins increases. Overall, increasing $M$ does
not provide a systematic improvement in the coverage of the shape
parameter and can introduce additional variability, particularly under
strong censoring and extreme values of $\xi$.

The results for the scale parameter $\sigma$
(Figure~\ref{fig:coverage_sigma}) show a similar qualitative pattern.
Increasing the number of bins generally leads to greater variability in
empirical coverage, with a tendency toward undercoverage for larger values
of $M$. The magnitude of this effect varies with the sample size, censoring
proportion, and censoring mechanism, but the general pattern is consistent
across the simulation scenarios.

The tail behaviour of the censoring distribution has a comparatively smaller
effect on coverage than the overall censoring proportion. Similar
qualitative patterns are observed under bounded, exponential, and
heavy-tailed censoring. This suggests that the proposed procedure is
reasonably robust to different censoring tail behaviours. In contrast, the
censoring proportion has a more pronounced effect on inferential
performance, particularly in the scenarios with heavier censoring.

Overall, the results indicate that the number of bins is an important tuning
parameter controlling the trade-off between discretization and statistical
variability. Smaller values of $M$, particularly $M=5$ and $M=10$, provide
reliable coverage across most of the scenarios considered, whereas larger
values can increase variability and lead to coverage deterioration.
Accordingly, we use
\(
M=10
\)
as the default specification in the subsequent simulation study. This
choice represents a compromise between maintaining sufficient resolution
in the piecewise approximation and avoiding excessive variability from
using a large number of intervals.

\subsection{Simulation study II: robustness to model misspecification}

The second simulation study evaluates the robustness of the proposed model
under distributional misspecification, that is, when the data-generating
distribution does not belong to the EGP family. Specifically, survival
times are generated from a lognormal distribution with parameters
$\mu=0$ and $\sigma=1$, and density
\begin{equation}
f(x;\mu,\sigma)
=
\frac{1}{x\sigma\sqrt{2\pi}}
\exp\left\{
-\frac{(\log x-\mu)^2}{2\sigma^2}
\right\},
\qquad
x>0.
\end{equation}

The censoring times are generated independently from the same lognormal
distribution. Because the survival and censoring times are independent and
identically distributed continuous random variables,
\(
\mathbb{P}(X\leq C)
=
\mathbb{P}(X>C)
=
\frac{1}{2},
\)
and hence the expected censoring proportion is $50\%$. As in the first simulation study, the observed follow-up times and
censoring indicators are defined by
\(
T_{i,j}
=
\min(X_{i,j},C_{i,j})
\) and \(
\delta_{i,j}
=
\mathds{1}\{X_{i,j}\leq C_{i,j}\}.
\)

Based on the results of the first simulation study, the number of bins is
fixed at
\(
M=10.
\)
In contrast to the first study, the parameter $\kappa$ is now treated as
unknown and is estimated jointly with the remaining model parameters.
Following the prior specification in
Section~\ref{priors_for_everyone}, we assign
\begin{equation}
\kappa\sim\operatorname{Lognormal}(0,10). 
\end{equation}

\begin{figure}[ht]
\centering
\includegraphics[width=\linewidth]{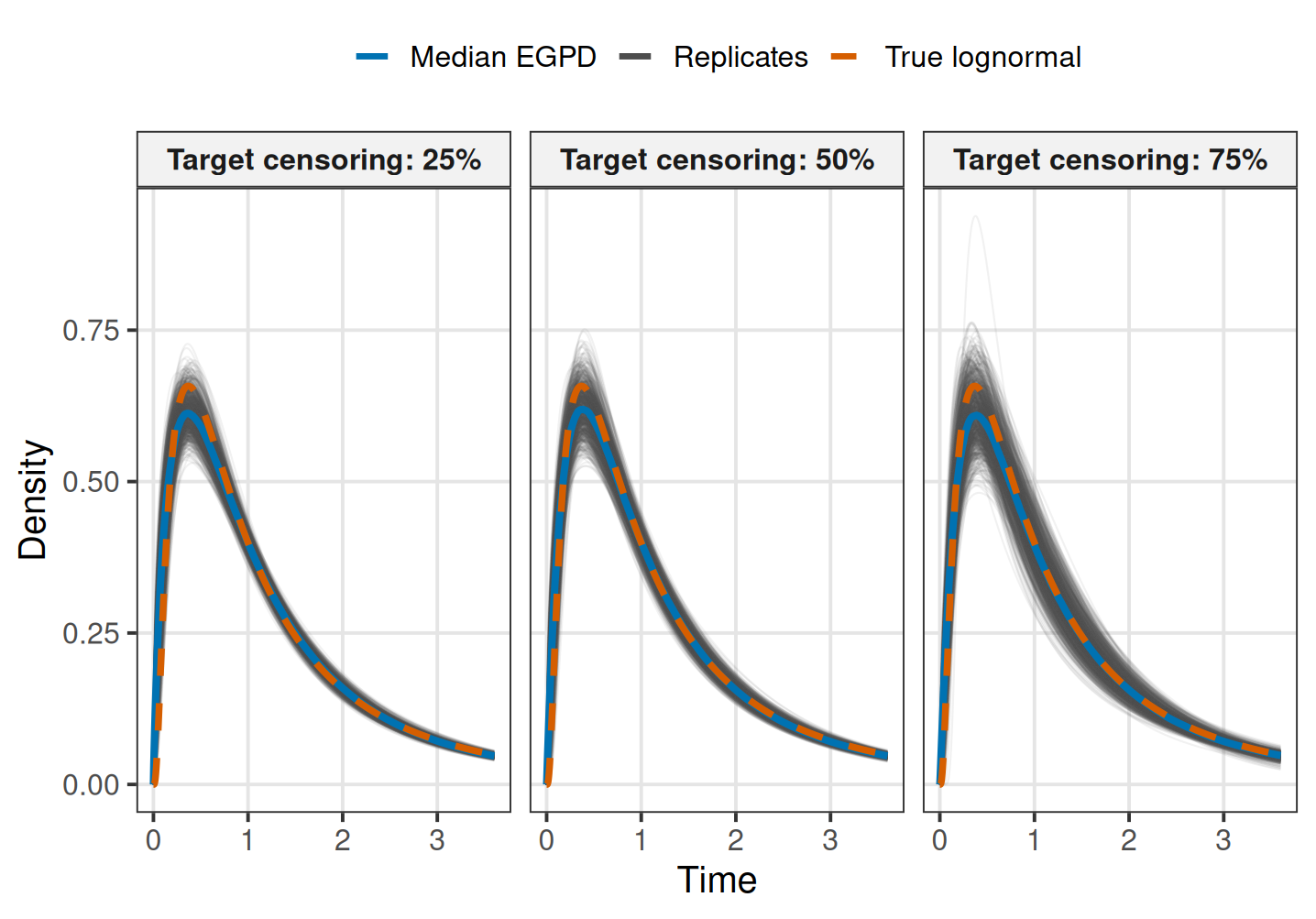}
\caption{Estimated densities obtained from the $1{,}000$ simulation
replicates under lognormal data generation. The grey curves correspond to
the estimated densities for individual replicates, the blue curve represents
the pointwise median estimated density, and the red curve is the true
lognormal density.}
\label{fig:lognormal_fit}
\end{figure}

For each configuration, $1{,}000$ independent simulation replicates are
generated. Average posterior summaries of the model parameters are reported
in Table~\ref{tab:lognormal_summary}. In addition, Figure~\ref{fig:lognormal_fit}
compares the true lognormal density with the densities estimated from the
simulation replicates. The individual estimated densities are displayed in
grey, the pointwise median estimated density is shown in blue, and the true
lognormal density is shown in red. This comparison provides a visual
assessment of the ability of the proposed model to approximate the
underlying distribution when the EGP assumption is misspecified.

\begin{table}[!h]
\centering
\caption{Average posterior summaries over the $1{,}000$ simulation
replicates under the lognormal data-generating mechanism, by target
censoring level.}
\label{tab:lognormal_summary}
\begin{tabular}{ccccccc}
\toprule
Cens. & Param. & Mean & Median & SD & 5\% Quantile & 95\% Quantile\\
\midrule
& $\kappa$ & 2.044 & 1.900 & 0.599 & 1.412 & 3.156\\
25\% & $\sigma$ & 0.691 & 0.693 & 0.156 & 0.430 & 0.943\\
& $\xi$ & 0.366 & 0.361 & 0.096 & 0.217 & 0.531\\
\cmidrule{1-7}
& $\kappa$ & 2.443 & 2.130 & 1.144 & 1.459 & 4.486\\
50\%& $\sigma$ & 0.621 & 0.621 & 0.185 & 0.315 & 0.926\\
& $\xi$ & 0.426 & 0.419 & 0.133 & 0.218 & 0.656\\
\cmidrule{1-7}
& $\kappa$ & 3.139 & 2.399 & 2.414 & 1.412 & 7.736\\
75\%& $\sigma$ & 0.605 & 0.598 & 0.250 & 0.207 & 1.026\\
& $\xi$ & 0.476 & 0.466 & 0.230 & 0.116 & 0.868\\
\bottomrule
\end{tabular}
\end{table}

Figure~\ref{fig:lognormal_fit} shows that the proposed model can provide a
close approximation to the underlying lognormal distribution despite the
fact that the data-generating mechanism does not belong to the EGP family.
Although the estimated densities exhibit variability across simulation
replicates, their overall shapes closely follow the true density. The
pointwise median estimated density is particularly close to the true
lognormal density, with a small discrepancy around the mode and some
redistribution of probability mass toward the lower tail.

This flexibility is partly enabled by the additional parameter $\kappa$,
which allows departures from the classical GP shape in the non-extreme
part of the distribution while retaining a GP-type tail representation.
The corresponding posterior summaries in
Table~\ref{tab:lognormal_summary} show that the estimated value of $\kappa$
increases as the censoring proportion increases, accompanied by increasing
posterior variability. The estimated values of $\xi$ are positive across
the three censoring levels, indicating that the fitted EGP representation
uses a heavy-tailed asymptotic component to approximate the lognormal
distribution over the range of the observed data.

It is important to note, however, that the lognormal distribution has
extreme value index $\xi=0$ in the asymptotic extreme-value sense. Thus,
the positive estimates of $\xi$ should not be interpreted as recovery of
the true asymptotic extreme value index. Rather, they reflect the ability
of the EGP model to approximate the lognormal tail over the finite,
non-asymptotic range represented in the data. Although the lognormal is
not regularly varying and therefore does not have a positive asymptotic
extreme value index, its tail is heavier than an exponential tail and can
exhibit substantial tail heaviness over finite ranges. Consequently, a
positive $\xi$ can provide a useful effective approximation to the
lognormal tail over the range of survival times accessible in the
simulation. The fact that the fitted model does not concentrate around
$\xi=0$ is therefore consistent with the distinction between finite-range
tail approximation and asymptotic tail behaviour.

These results should consequently be interpreted as an assessment of
approximation robustness rather than as evidence that the lognormal
distribution belongs to the EGP family. In particular, the positive
estimates of $\xi$ illustrate that the fitted EGP representation can use
its heavy-tailed component to capture the effective tail behaviour of a
lognormal distribution over the observed and extrapolated range, even
though its asymptotic extreme value index is zero. Overall, the simulation
provides evidence that the proposed intensity-based estimation procedure
can retain useful distributional flexibility under moderate model
misspecification.

\section{Application}
\label{sec:application}

To illustrate the proposed methodology, we analyze a real-world dataset consisting of patients diagnosed with heart failure and followed at the Heart Institute of the University of S\~ao Paulo Medical School (INCOR--FMUSP), Brazil \cite{lima2015analise}. Heart failure is a complex clinical syndrome in which the heart is unable to pump sufficient blood to meet the body's metabolic demands and represents a major cause of morbidity and mortality worldwide.

The original cohort comprised patients who entered the study between July 2003 and March 2014 and were subsequently monitored through regular clinical follow-up. Patients were followed until death or the end of the observation period, with individuals who remained alive or were lost to follow-up treated as right-censored. After data cleaning and consistency checks, the final dataset consisted of 2,982 patients, of whom 1,286 experienced the event of interest (all-cause mortality) and 1,696 (56.87\%) were right-censored.

For each patient, the observed survival time was defined as the elapsed time, measured in months, from study entry to either the date of death or the date of the last available follow-up. The analysis uses three variables: follow-up time, the event indicator, and sex. The primary objective is to compare the survival experience of male and female patients using the proposed EGP model. The study population consists of 1,705 male patients (791 deaths and 914 censored observations) and 1,277 female patients (495 deaths and 782 censored observations).

To incorporate the effect of sex, male patients are taken as the reference category. Let $Z=0$ denote the male group and $Z=1$ the female group. Sex is allowed to affect both the scale and tail parameters, while the EGP parameter $\kappa$ is assumed to be common to both groups. This choice reflects the clinical assumption that survival patterns over the very early observation window are broadly similar between males and females, before differences in longer-term survival and tail behaviour become apparent. The shared parameter $\kappa$ captures this initial survival pattern, while sex-specific scale and tail parameters allow the groups to differ in their overall survival experience and long-term risk.
\begin{equation}
\sigma(Z)=\exp(\sigma_0+\sigma_1 Z),
\qquad
\xi(Z)=\beta_0+\beta_1 Z,
\qquad
\kappa=\exp(\kappa_0).
\end{equation}

For a given value of $Z$, define
\begin{equation}
\label{eq:q_z}
q_Z(t)
=
\begin{cases}
\displaystyle
\left(
1+\frac{\xi(Z)t}{\sigma(Z)}
\right)^{-1/\xi(Z)},
& \xi(Z)\neq0,\\[10pt]
\displaystyle
\exp\left\{-\frac{t}{\sigma(Z)}\right\},
& \xi(Z)=0.
\end{cases}
\end{equation}

The EGP hazard is then given by
\begin{equation}
\label{eq:egp_hazard_sex}
h(t\mid Z)
=
\frac{\kappa}{\sigma(Z)}
\frac{
[1-q_Z(t)]^{\kappa-1}
q_Z(t)^{1+\xi(Z)}
}{
1-[1-q_Z(t)]^\kappa
}.
\end{equation}

For the male group ($Z=0$), the hazard reduces to
\begin{equation}
\label{eq:male_hazard}
h(t\mid Z=0)=h_0(t)
=
\frac{e^{\kappa_0}}{e^{\sigma_0}}
\frac{
[1-q_0(t)]^{e^{\kappa_0}-1}
q_0(t)^{1+\beta_0}
}{
1-[1-q_0(t)]^{e^{\kappa_0}}
},
\end{equation}
where
\begin{equation}
\label{eq:q0}
q_0(t)
=
\begin{cases}
\displaystyle
\left(
1+\frac{\beta_0 t}{e^{\sigma_0}}
\right)^{-1/\beta_0},
& \beta_0\neq0,\\[8pt]
\displaystyle
\exp\left(-\frac{t}{e^{\sigma_0}}\right),
& \beta_0=0.
\end{cases}
\end{equation}

For the female group ($Z=1$), the corresponding hazard is
\begin{equation}
\label{eq:female_hazard}
h(t\mid Z=1)=h_1(t)
=
\frac{e^{\kappa_0}}{e^{\sigma_0+\sigma_1}}
\frac{
[1-q_1(t)]^{e^{\kappa_0}-1}
q_1(t)^{1+\beta_0+\beta_1}
}{
1-[1-q_1(t)]^{e^{\kappa_0}}
},
\end{equation}
where
\begin{equation}
\label{eq:q1}
q_1(t)
=
\begin{cases}
\displaystyle
\left(
1+
\frac{(\beta_0+\beta_1)t}
{e^{\sigma_0+\sigma_1}}
\right)^{-1/(\beta_0+\beta_1)},
& \beta_0+\beta_1\neq0,\\[8pt]
\displaystyle
\exp\left(-\frac{t}{e^{\sigma_0+\sigma_1}}\right),
& \beta_0+\beta_1=0.
\end{cases}
\end{equation}

The Bayesian inference procedure and prior distributions follow
Sections~\ref{sec:MYMODEL} and~\ref{sec:LookAtMyTail}, respectively,
with the additional regression structure for the scale and tail
parameters. On the transformed scale, the regression parameters have
independent Gaussian priors:
\begin{equation}
\sigma_0,\sigma_1,\kappa_0
\stackrel{\mathrm{ind}}{\sim}
\mathcal{N}(0,10).
\end{equation}
The coefficients $\beta_0$ and $\beta_1$ follow the sequential prior
construction described in Section~\ref{sec:LookAtMyTail}, which
preserves the admissibility of the GP support over the observed time
domain.

\begin{figure}[ht]
    \centering
    \includegraphics[width=0.8\textwidth]{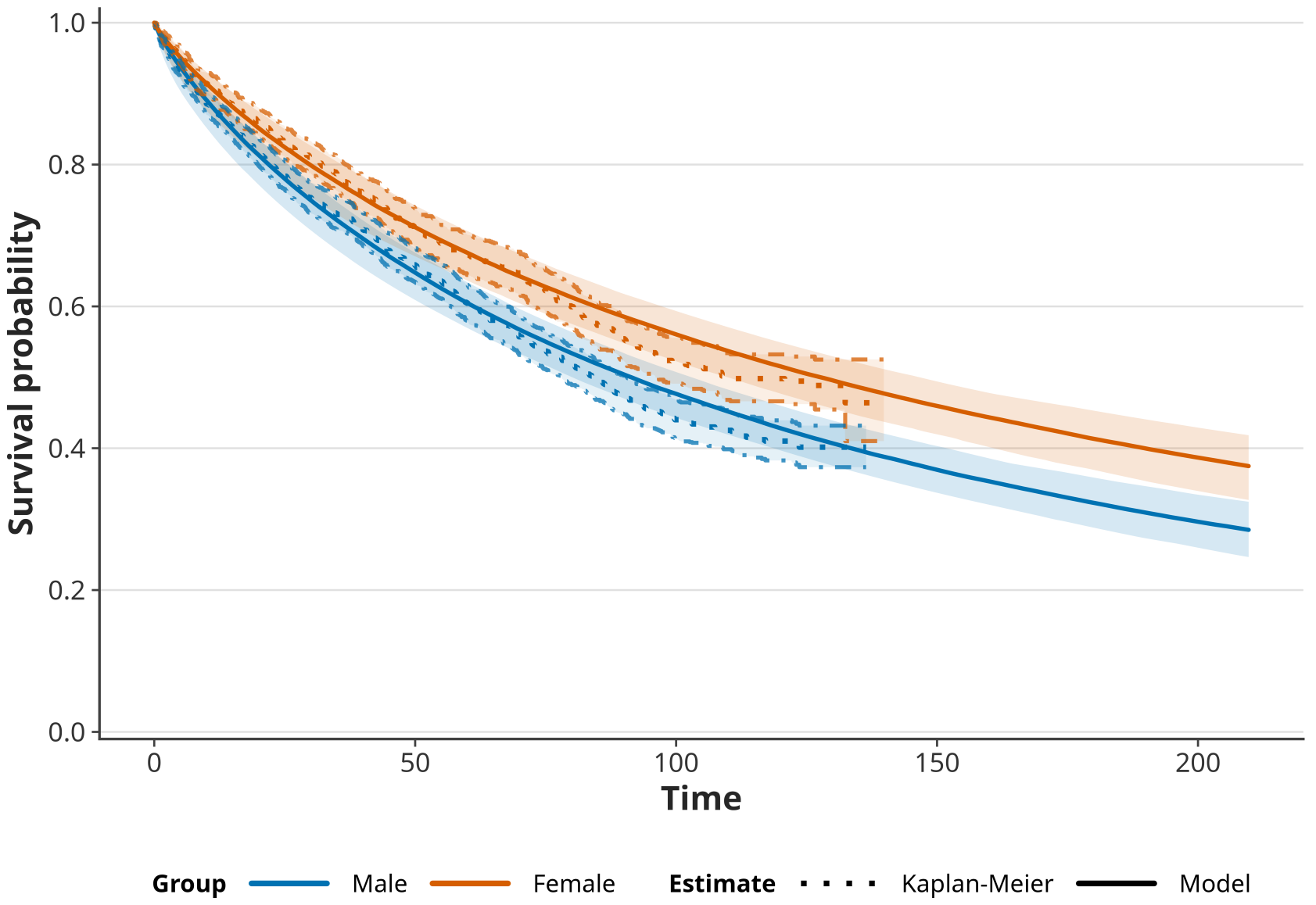}
    \caption{Posterior median survival curves with 95\% credible intervals (shaded ribbons) compared with the Kaplan--Meier estimates and their 95\% confidence intervals, stratified by sex.}
    \label{fig:survival_km_comparison}
\end{figure}

Table~\ref{tab:model_summary} presents the posterior summaries of the
model parameters, while Figure~\ref{fig:survival_km_comparison}
compares the posterior survival curves with the corresponding
Kaplan--Meier estimates for male and female patients.

\begin{table}[htbp]
\centering
\caption{Posterior summaries of the model parameters.}
\label{tab:model_summary}
\begin{tabular}{cccccc}
\toprule
Parameter & Mean & SD & 5\% & 50\% & 95\% \\
\midrule
$\beta_0$   & 0.63  & 0.26 & 0.20  & 0.63 & 1.04 \\
$\beta_1$   & 0.11  & 0.35 & -0.46 & 0.10 & 0.70 \\
$\sigma_0$  & 4.89  & 0.19 & 4.60  & 4.87 & 5.22 \\
$\sigma_1$  & 0.31  & 0.11 & 0.13  & 0.30 & 0.49 \\
$\kappa_0$  & -0.18 & 0.11 & -0.39 & -0.16 & -0.04 \\
\bottomrule
\end{tabular}
\end{table}

The posterior distributions of the regression coefficients associated with the female group provide different levels of evidence across the model parameters. The 90\% credible interval for the tail parameter $\beta_1$ contains zero, indicating limited evidence that sex affects the extreme value index. Thus, although the estimated posterior median is positive, the data do not provide sufficient evidence of a systematic difference in tail heaviness between male and female patients.

In contrast, the posterior distribution of the scale coefficient $\sigma_1$ is entirely positive, with its 90\% credible interval excluding zero. This provides evidence of a sex difference in the scale of the survival distribution. Specifically, since the scale is modelled on the logarithmic scale, $\sigma_1>0$ implies
\begin{equation}
\sigma(1)=\exp(\sigma_0+\sigma_1)
>
\exp(\sigma_0)=\sigma(0).
\end{equation}
Thus, the results support a difference in the overall survival time scale between females and males, but do not provide equally strong evidence of a sex difference in the asymptotic tail behaviour.

The posterior median survival functions closely follow the corresponding Kaplan--Meier estimates for both sexes, as shown in Figure~\ref{fig:survival_km_comparison}, indicating an adequate fit to the observed data. Although the estimated survival curves remain broadly similar at longer follow-up times, the positive posterior effect on the scale parameter indicates a systematic difference in the time scale of survival between the two groups. At the same time, the absence of strong evidence for $\beta_1\neq0$ suggests that the asymptotic tail behaviour, as characterized by the extreme value index, is not substantially different between male and female patients.

\section{Conclusions}
\label{sec:conclusions}

This paper introduced a Bayesian framework for long-term survival inference under right censoring based on an intensity representation of the generalized Pareto and extended generalized Pareto distributions. The main contribution is an alternative estimation procedure for GP- and EGP-based survival models that exploits a piecewise exponential representation of the hazard and a latent Poisson-process formulation. This construction avoids the need for direct closed-form likelihood evaluation while retaining the extreme-value structure required for extrapolation beyond the observed follow-up period.

A central advantage of the proposed formulation is that the extreme value index $\xi$ can be estimated without imposing an a priori restriction to a specific tail regime. In particular, the parameterization allows the data to distinguish between finite-endpoint ($\xi<0$), exponential ($\xi=0$), and heavy-tailed ($\xi>0$) behaviour, subject only to the support constraints of the specified time domain. This feature is particularly relevant in survival analysis, where assuming a heavy-tailed distribution in advance may be difficult to justify. The use of the full set of observed follow-up times also avoids the need to select a conventional POT threshold separating bulk and extreme observations.

The proposed two-stage procedure first approximates the continuous hazard by a piecewise exponential model and subsequently treats the resulting interval-specific hazard estimates as noisy observations of the underlying GP or EGP hazard. Latent evaluation locations are introduced to account for the fact that an interval-level hazard estimate does not correspond naturally to a unique deterministic time point. Bayesian inference is then performed using HMC, providing posterior distributions for the scale, tail, and EGP shape parameters and consequently for the extrapolated survival function.

The simulation studies indicate that the proposed procedure provides reliable posterior uncertainty quantification across a broad range of tail regimes, censoring proportions, censoring distributions, and sample sizes. The first simulation study showed that a moderate number of intervals, particularly $M=5$ or $M=10$, generally provides satisfactory empirical coverage, with $M=10$ adopted as the default specification in subsequent experiments. Importantly, the procedure was evaluated for negative, zero, and positive values of $\xi$, demonstrating that its performance does not depend on imposing a predetermined short- or long-tail assumption. The second simulation study further showed substantial robustness to model misspecification: when the data were generated from a lognormal distribution, the EGP model was able to recover the overall distributional shape despite the data-generating mechanism lying outside the assumed model family.

The application to the heart failure dataset demonstrated how the proposed
framework can be used to investigate long-term survival under substantial
right censoring. The inclusion of covariate effects on both the scale and
tail parameters allows differences between groups to be examined separately
in terms of overall survival levels and asymptotic tail behaviour. In the
application, the posterior results provided evidence for a difference in the
scale parameter between male and female patients, while the posterior
credible interval for the tail effect included zero, suggesting limited
evidence of a difference in the asymptotic tail behaviour between the two
groups.

The bias-correction mechanisms developed in this work provide additional
tools for addressing discretization effects arising from the
piecewise-exponential approximation. In particular, latent
bias-correcting representative locations and adaptive partition refinement
can be used to reduce curvature and tail-amplification effects. 
Nevertheless, they provide a natural direction for improving the procedure
in settings where the tail is particularly heavy or where the available
sample size is limited.

Several directions for future research remain. A systematic investigation of
the adaptive partitioning procedure and its interaction with posterior
inference would be particularly valuable, as would principled data-driven
selection of the number of intervals and the concentration parameter
governing the latent representative locations. Further extensions could consider competing risks, recurrent
events, multistate processes, and dependent censoring, thereby broadening
the applicability of the proposed intensity-based framework.

Overall, the proposed methodology provides a flexible and computationally
tractable alternative for GP and EGP estimation under right censoring.
By replacing direct likelihood-based estimation with an intensity-based
piecewise Poisson representation, the framework facilitates Bayesian
inference for the extreme value index without requiring a priori
classification of the tail as short, exponential, or heavy. This combination
of extreme-value structure, censoring-aware survival modelling, and Bayesian
regularization provides a useful framework for estimating and extrapolating
long-term survival probabilities when observations in the extreme tail are
sparse or heavily censored.

\section*{Acknowledgments}
Eduardo Janotti is supported by the São Paulo Research Foundation (FAPESP), Brazil, under Grant Nos.\ 2025/10342-3 and 2025/22925-3. Lígia Henriques-Rodrigues and Eduardo janotti were funded by national funds through FCT – Fundação para a Ciência e a Tecnologia, I.P., under CEAUL Research Unit, UID/00006/2025, \url{DOI: https://doi.org/10.54499/UID/00006/2025}, and by the European Union – NextGenerationEU through the project UID/PRR/00006/2025, \url{DOI: https://doi.org/10.54499/UID/PRR/00006/2025.}

\bibliographystyle{tfp}
\bibliography{bibliografia.bib}

\newpage
\appendix

\section{Convergence of the Piecewise Exponential Approximation}
\label{app:convergence}

This appendix establishes the asymptotic validity of the piecewise
exponential approximation used in the proposed two-stage inference
procedure. The argument is based on two sources of approximation error:
the statistical error associated with estimating the interval-specific
hazards and the discretization error arising from replacing the continuous
hazard by a piecewise constant function.

Throughout this appendix, we assume that the tail distribution belongs to
the generalized Pareto family. This assumption is motivated by the
Pickands--Balkema--de Haan theorem, under which the conditional excess
distribution above a sufficiently high threshold is asymptotically
approximated by a GP. Accordingly, the asymptotic hazard in the tail may
be represented by the GP hazard.

Let
\[
T\sim\mathrm{GP}(\sigma,\xi),
\qquad
\sigma>0,
\]
with hazard function
\[
\lambda(t)
=
\frac{1}{\sigma+\xi t},
\]
on its support
\[
1+\frac{\xi t}{\sigma}>0.
\]
For $\xi<0$, the support has a finite upper endpoint
$-\sigma/\xi$; for $\xi\geq0$, the support is unbounded.

Let $\mathcal{T}=[0,T]$ be a fixed compact time interval contained in the
support of the distribution, and consider a sequence of partitions
\[
0=\rho_1<\rho_2<\cdots<\rho_{M+1}=T,
\]
with intervals
\[
I_j=(\rho_j,\rho_{j+1}),
\qquad j=1,\ldots,M.
\]
Denote the maximal mesh width by
\[
\Delta_M
=
\max_{1\leq j\leq M}
(\rho_{j+1}-\rho_j).
\]
We consider the asymptotic regime
\[
n\to\infty,
\qquad
M\to\infty,
\qquad
\Delta_M\to0,
\]
with the number of observed events in each interval increasing with the
sample size.

Let $N(t)$ denote the counting process of observed events and let
$Y(t)$ denote the corresponding at-risk process. For interval $I_j$, define
the number of observed events and the total exposure time by
\[
D_j
=
\int_{I_j}dN(t),
\qquad
E_j
=
\int_{I_j}Y(t)\,dt.
\]
The piecewise exponential estimator of the hazard is
\[
\widehat{\lambda}_j
=
\frac{D_j}{E_j}.
\]

Under independent censoring and the usual regularity conditions for
counting-process estimators, the compensator of $N(t)$ is
\[
\Lambda_N(t)
=
\int_0^tY(u)\lambda(u)\,du.
\]
Consequently, the law of large numbers for counting processes implies
that, for each fixed interval $I_j$,
\[
\widehat{\lambda}_j
\xrightarrow{p}
\bar{\lambda}_j,
\]
where
\[
\bar{\lambda}_j
=
\frac{
\int_{I_j}Y(t)\lambda(t)\,dt
}{
\int_{I_j}Y(t)\,dt
}
\]
is the exposure-weighted average hazard over the interval.

The quantity $\bar{\lambda}_j$ need not equal the hazard at the midpoint
of the interval. However, because the GP hazard is continuously
differentiable on every compact subset of its support, the discrepancy
between the interval average and the hazard evaluated at any point in the
interval becomes arbitrarily small as the interval width decreases.

Indeed, for $t\in\mathcal{T}$,
\[
\lambda'(t)
=
-\frac{\xi}{(\sigma+\xi t)^2}.
\]
Since $\mathcal{T}$ is compact and lies in the interior of the support,
there exists a finite constant $C_T>0$ such that
\[
\sup_{t\in\mathcal{T}}|\lambda'(t)|
\leq C_T.
\]
Therefore, for any $s,t\in I_j$,
\[
|\lambda(s)-\lambda(t)|
\leq
C_T|s-t|
\leq
C_T\Delta_M.
\]
Since $\bar{\lambda}_j$ is a weighted average of $\lambda(t)$ over
$I_j$, it follows that, for any $t_j\in I_j$,
\[
|\bar{\lambda}_j-\lambda(t_j)|
\leq
C_T\Delta_M.
\]
Hence,
\[
\max_{1\leq j\leq M}
|\bar{\lambda}_j-\lambda(t_j)|
\leq
C_T\Delta_M
\longrightarrow0.
\]

Thus, the discretization error vanishes as the maximum interval width tends
to zero. Combining this result with the consistency of the interval hazard
estimators yields
\[
\widehat{\lambda}_j
-
\lambda(t_j)
=
\left(
\widehat{\lambda}_j-\bar{\lambda}_j
\right)
+
\left(
\bar{\lambda}_j-\lambda(t_j)
\right)
\xrightarrow{p}0,
\]
provided that the number of events and the corresponding exposure in each
interval are sufficiently large.

\begin{proposition}[Consistency of the piecewise exponential approximation]
\label{prop:piecewise-consistency}
Assume that the true hazard is the hazard of a GP on a compact interval
$\mathcal{T}=[0,T]$ contained in the interior of its support. Suppose that
censoring is independent and that the counting-process regularity
conditions required for consistency of the Nelson--Aalen type interval
hazard estimator hold. Let the partition satisfy
\[
\Delta_M\to0
\]
and suppose that the number of observed events and the exposure time in
each interval increase sufficiently rapidly with $n$. Then, for any
sequence of representative points $t_j\in I_j$,
\[
\max_{1\leq j\leq M}
\left|
\widehat{\lambda}_j-\lambda(t_j)
\right|
\xrightarrow{p}0.
\]
\end{proposition}

\begin{proof}
The interval estimator converges to the exposure-weighted average hazard,
\[
\widehat{\lambda}_j-\bar{\lambda}_j
\xrightarrow{p}0.
\]
Moreover, the differentiability of the GP hazard on the compact interval
implies
\[
\max_j
|\bar{\lambda}_j-\lambda(t_j)|
\leq C_T\Delta_M
\longrightarrow0.
\]
The result follows from the triangle inequality.
\end{proof}

The latent evaluation locations introduced in the main text can therefore
be interpreted as a probabilistic representation of the fact that an
interval-specific hazard estimate corresponds to an interval rather than
to a unique deterministic time point. In particular, if
\[
L_j\in I_j,
\]
then the Lipschitz property of the hazard gives
\[
|\lambda(L_j)-\lambda(t_j)|
\leq
C_T\Delta_M.
\]
Consequently,
\[
\max_j
|\lambda(L_j)-\lambda(t_j)|
\xrightarrow{}0.
\]
Combining this result with Proposition~\ref{prop:piecewise-consistency}
gives
\[
\max_j
|\widehat{\lambda}_j-\lambda(L_j)|
\xrightarrow{p}0.
\]

Importantly, this result does not require $L_j$ to be uniformly
distributed. The asymptotic argument only requires
$L_j\in I_j$. The uniform distribution adopted in the main model is a
modelling choice used to represent uncertainty about the location of the
interval-specific hazard and is not required for consistency.

\subsection{Extension to the EGP hazard}

The same argument applies to the EGP hazard provided that its parameters
remain in the interior of the admissible parameter space and the resulting
hazard is continuously differentiable on the compact time interval under
consideration.

Recall that the EGP hazard is
\[
h_{\mathrm{EGP}}(t;\sigma,\xi,\kappa)
=
\frac{\kappa}{\sigma}
\frac{
[1-q(t)]^{\kappa-1}q(t)^{1+\xi}
}{
1-[1-q(t)]^\kappa
},
\]
where
\[
q(t)
=
\left(
1+\frac{\xi t}{\sigma}
\right)^{-1/\xi}.
\]
For any compact interval contained in the interior of the support and for
admissible parameter values, $h_{\mathrm{EGP}}$ is continuous and
continuously differentiable. Therefore, the same Lipschitz argument yields
\[
\max_j
\left|
\bar h_j-h_{\mathrm{EGP}}(t_j;\sigma,\xi,\kappa)
\right|
\leq
C\Delta_M
\longrightarrow0,
\]
for some finite constant $C$.

Hence, the piecewise exponential approximation is asymptotically valid
for both the classical GP hazard and its EGP extension.

\subsection{Remarks on the infinite horizon}

The preceding results are established on compact subsets of the support.
This restriction is important and should not be interpreted as implying
uniform convergence of the hazard on the entire half-line.

For $\xi\geq0$, the GP support is unbounded, but the hazard may become
arbitrarily small as $t$ increases. For $\xi<0$, the distribution has a
finite upper endpoint
\[
t_F=-\frac{\sigma}{\xi},
\]
and the hazard diverges as $t$ approaches this endpoint. Consequently,
uniform approximation over the entire support cannot in general be obtained
from the compact-interval argument alone.

For any fixed $\varepsilon>0$, however, one may restrict attention to a
compact interval
\[
\mathcal{T}_\varepsilon=[0,\rho_\varepsilon]
\]
contained in the interior of the support. The preceding results then imply
that
\[
\sup_{t\in\mathcal{T}_\varepsilon}
|\widehat{\lambda}_M(t)-\lambda(t)|
\xrightarrow{p}0
\]
as the sample size increases and the maximum partition width decreases.

For $\xi\geq0$, because the GP survival function satisfies
\[
S(t)\to0
\qquad\text{as }t\to\infty,
\]
$\rho_\varepsilon$ can additionally be chosen so that
\[
S(\rho_\varepsilon)<\varepsilon.
\]
Thus, the probability of observing an event beyond
$\rho_\varepsilon$ is smaller than $\varepsilon$. This provides a probabilistic
justification for restricting the approximation to a sufficiently large
finite time horizon when the objective is survival inference over the
observationally relevant region.

For $\xi<0$, the corresponding restriction is naturally imposed below the
finite endpoint $t_F$. In this case, compact subsets bounded away from
$t_F$ are required for the uniform convergence argument.

\begin{proposition}[Compact-set convergence]
\label{prop:compact-convergence}
Under the assumptions of Proposition~\ref{prop:piecewise-consistency}, let
$\mathcal{K}$ be any compact subset of the interior of the GP or EGP
support. Then the piecewise exponential approximation converges uniformly
in probability to the corresponding true hazard on $\mathcal{K}$:
\[
\sup_{t\in\mathcal{K}}
\left|
\widehat{\lambda}_M(t)-\lambda(t)
\right|
\xrightarrow{p}0.
\]
\end{proposition}

Therefore, the proposed two-stage procedure is asymptotically coherent on
every compact subset of the admissible time domain. The approximation error
has two components: statistical estimation error, which vanishes as the
sample size and interval exposure increase, and discretization error, which
vanishes as the maximum partition width tends to zero. This provides the
theoretical justification for using the piecewise exponential
representation as an approximation to the continuous GP or EGP hazard in
the proposed Bayesian inference procedure.

\section{Convergence analysis of the bias correction procedure}

This appendix studies the effect of interval discretization on the estimation
of the extreme value index and provides theoretical justification for the
proposed representative-point correction. For simplicity, the analysis is
conducted for the standard GP, whose
hazard function is
$$
h(x)=\frac{1}{\sigma+\xi x}.
$$

The case $\xi=0$ corresponds to the exponential distribution, for which the
hazard is constant and discretization therefore introduces no bias. We
henceforth restrict attention to $\xi\neq0$.

The analysis relies on the stability of the GP under threshold
exceedances. In particular, let $\tau$ be a threshold satisfying

$$
1+\frac{\xi\tau}{\sigma}>0.
$$

Using the GP survival function,

$$
\overline F(x)
=
\mathbb P(X>x)
=
\left(
1+\frac{\xi x}{\sigma}
\right)^{-1/\xi},
$$

the survival function of the excess variable

$$
Y=X-\tau\mid X>\tau
$$

is

$$
\begin{aligned}
\mathbb P(Y>y)
&=
\mathbb P(X>\tau+y\mid X>\tau)\\
&=
\frac{\overline F(\tau+y)}{\overline F(\tau)}\\
&=
\left[
\frac{
1+\frac{\xi(\tau+y)}{\sigma}
}{
1+\frac{\xi\tau}{\sigma}
}
\right]^{-1/\xi}\\
&=
\left(
1+\frac{\xi y}{\sigma+\xi\tau}
\right)^{-1/\xi}.
\end{aligned}
$$

Therefore,

$$
X-\tau\mid X>\tau
\sim
\operatorname{GP}(\sigma+\xi\tau,\xi).
$$

Thus, thresholding changes only the scale parameter, from $\sigma$ to

$$
\sigma_\tau=\sigma+\xi\tau,
$$

while preserving the extreme value index $\xi$. This result allows the
discretization analysis to be transferred from an interval beginning at
zero to any subsequent interval. Specifically, conditional on exceeding
the lower endpoint $\ell_j$ of an interval, the excess time

$$
X-\ell_j\mid X>\ell_j
$$

follows a GP with scale

$$
\sigma_j=\sigma+\xi\ell_j
$$

and the same shape parameter $\xi$. Consequently, the representative-point
analysis can be carried out without loss of generality on a first interval
of the form

$$
I_1=[0,L),
$$

with the results for subsequent intervals obtained by replacing $\sigma$
with the corresponding threshold-adjusted scale $\sigma_j$.

\subsection{Representative point associated with the interval-average hazard}

Consider the interval
\[
I_1=[0,L).
\]

The average hazard over this interval is
\[
\bar h
=
\frac{1}{L}
\int_0^L
\frac{1}{\sigma+\xi t}
\,dt.
\]

Evaluating the integral yields
\[
\bar h
=
\frac{1}{\xi L}
\log
\left(
1+\frac{\xi L}{\sigma}
\right).
\]

Define the representative point $r_1$ as the location satisfying
\[
h(r_1)=\bar h.
\]

Substituting the GP hazard gives
\[
\frac{1}{\sigma+\xi r_1}
=
\frac{1}{\xi L}
\log
\left(
1+\frac{\xi L}{\sigma}
\right).
\]

Solving for $r_1$ yields
\[
r_1
=
\frac{L}
{
\log\!\left(1+\frac{\xi L}{\sigma}\right)
}
-
\frac{\sigma}{\xi}.
\]

The conventional midpoint approximation instead evaluates the hazard at
\[
x_1=\frac{L}{2}.
\]

The discrepancy between these two evaluation points determines the discretization bias.

\subsection{Comparison with midpoint evaluation}

Introduce the ratio
\[
R(L,\sigma,\xi)
=
\frac{r_1}{x_1}.
\]

Substituting the expressions above gives
\[
R(L,\sigma,\xi)
=
2
\left(
\frac{1}
{
\log\!\left(1+\frac{\xi L}{\sigma}\right)
}
-
\frac{\sigma}{\xi L}
\right).
\]

Since the ratio depends only on the dimensionless quantity
\[
z=\frac{\xi L}{\sigma},
\]
we may write
\[
R(z)
=
2
\left(
\frac{1}{\log(1+z)}
-
\frac{1}{z}
\right).
\]

Using the expansion
\[
\log(1+z)
=
z-\frac{z^2}{2}+O(z^3),
\]
it follows that
\[
R(z)
=
1-\frac{z}{6}+O(z^2).
\]

Therefore,
\[
\lim_{z\to0}R(z)=1,
\]
showing that the midpoint and representative point coincide asymptotically as the interval width tends to zero.

Differentiating gives
\[
R'(z)
=
2
\left(
\frac{1}{z^2}
-
\frac{1}
{
(1+z)\log^2(1+z)
}
\right).
\]

For $z>0$,
\[
R'(z)<0,
\]
and therefore $R(z)$ is strictly decreasing. Since $R(0)=1$, we obtain
\[
R(z)<1,
\qquad z>0,
\]
and
\[
R(z)>1,
\qquad z<0.
\]

Consequently,
\[
\xi>0
\quad\Longrightarrow\quad
r_1<x_1,
\]
while
\[
\xi<0
\quad\Longrightarrow\quad
r_1>x_1.
\]

The discrepancy increases as $|\xi|L/\sigma$ grows.

\subsection{Attenuation of the extreme value index}

Suppose that the interval-average hazard is replaced by midpoint evaluation. Let $\xi'$ denote the parameter obtained by matching
\[
h(r_1;\sigma,\xi)
=
h(x_1;\sigma,\xi').
\]

Using the GP hazard,
\[
\frac{1}{\sigma+\xi r_1}
=
\frac{1}{\sigma+\xi' x_1}.
\]

Therefore,
\[
\xi'
=
\xi\frac{r_1}{x_1}
=
\xi R(L,\sigma,\xi).
\]

From the behaviour of $R(z)$, it follows that
\[
\xi'<\xi,
\qquad
\xi\neq0.
\]

Thus, midpoint evaluation systematically underestimates the extreme value index, resulting in a shorter estimated tail. This bias arises because the midpoint $x_1=L/2$ does not coincide with the representative point associated with the interval-average hazard. As a consequence, replacing the average hazard within an interval by its value at the midpoint leads to a distorted characterization of the tail behaviour.

The proposed correction instead evaluates the hazard at a representative point $r(\xi)$ satisfying
\[
h(r(\xi))
=
\frac{1}{L}
\int_0^L h(t)\,dt.
\]

By construction, the hazard evaluated at $r(\xi)$ coincides exactly with the interval-average hazard. The discrepancy introduced by midpoint evaluation is therefore removed, yielding a representative point that correctly reflects the average behaviour of the hazard within the interval.

In summary, midpoint discretization introduces a systematic negative bias in the estimated extreme value index. For heavy-tailed distributions ($\xi>0$), this results in underestimation of tail heaviness and consequently shorter estimated tails. The proposed representative-point correction eliminates this source of bias by aligning the evaluation location with the interval-average hazard, thereby providing a more faithful approximation of the underlying hazard structure.

\end{document}